\documentclass[fleqn,usenatbib]{mnras}

\usepackage{newtxtext,newtxmath}

\usepackage[T1]{fontenc}
\usepackage{ae,aecompl}
\usepackage{cancel}

\usepackage{graphicx}	
\usepackage{amsmath}	
\usepackage{amssymb}	
\usepackage{url}

\newcommand{\pderiv}[2]{\ensuremath{\frac{\partial #1}{\partial #2}}}

\title[Chemistry \& transport in discs]{Planet-forming material in a protoplanetary disc: the interplay between chemical evolution and pebble drift} 

\author[Booth \& Ilee]{R.~A.~Booth$^{1}$\thanks{E-mail: \href{rab200@ast.cam.ac.uk}{rab200@ast.cam.ac.uk} (RAB)}
and J.~D.~Ilee$^{2}$\thanks{E-mail: \href{j.d.ilee@leeds.ac.uk}{j.d.ilee@leeds.ac.uk} (JDI)}
\vspace{0.4em}
\\
$^{1}$Institute of Astronomy, Madingley Road, Cambridge CB3 0HA, UK\\
$^{2}$School of Physics \& Astronomy, University of Leeds, Leeds LS2 9JT, UK
}

\date{Accepted 2019 May 27. Received 2019 May 17; in original form 2019 April 10}

\pubyear{2019}

\begin{document}
\label{firstpage}
\pagerange{\pageref{firstpage}--\pageref{lastpage}}
\maketitle

\begin{abstract}
The composition of gas and solids in protoplanetary discs sets the composition of planets that form out of them. Recent chemical models have shown that the composition of gas and dust in discs evolves on Myr time-scales, with volatile species disappearing from the gas phase. However, discs evolve due to gas accretion and radial drift of dust on time-scales similar to these chemical time-scales. Here we present the first model coupling the chemical evolution in the disc mid-planes with the evolution of discs due to accretion and radial drift of dust. Our models show that transport will always overcome the depletion of CO$_2$ from the gas phase, and can also overcome the depletion of CO and CH$_4$ unless both transport is slow (viscous $\alpha \lesssim 10^{-3}$) and the ionization rate is high ($\zeta \approx 10^{-17}$). Including radial drift further enhances the abundances of volatile species because they are carried in on the surface of grains before evaporating left at their ice lines. Due to large differences in the abundances within 10~au for models with and without efficient radial drift, we argue that composition can be used to constrain models of planet formation via pebble accretion.
\end{abstract}




\section{Introduction}

The structure and composition of protoplanetary discs are fundamental pieces in the puzzle of how planets form and evolve. In the most direct sense, planets form via the accretion of gas and solids -- in the form of planetesimals, pebbles, or dust, which together with the gas are the main components of protoplanetary discs. This provides the opportunity to learn about planet formation by connecting the composition of planets to the discs in which they form. A challenge for the protoplanetary disc community is thus to determine the composition of the planetary building blocks, a task complicated by uncertainties in the processes governing disc evolution and the associated difficulties with the interpretation observational constraints. Here, we explore one aspect of this problem -- namely how differences in the transport of gas and solids through the disc compete with their chemical evolution to control the composition of discs. 

\smallskip

We frame our investigation in terms of a recent idea, which is to use the atmospheric C/O ratio in hot Jupiters to determine their building blocks. The observational utility of the C/O ratio comes from the relative ease with which it can be constrained in hot Jupiter atmospheres.  This is because the abundance of the observed molecules can vary by orders of magnitude for only  small changes in the C/O ratio \citep[see][]{lodders_2010, madhusudhan_2012}. \citet{oberg_2011} realised that the C/O ratio could be used to constrain the radial location within a disc from which a planet formed.  This is possible because the C/O ratio of the gas and ices changes across the ice lines (places where molecular species transition from being predominantly in the gas phase to ices in the solid phase) of the dominant carbon and oxygen bearing species, such as CO, CO$_2$ and H$_2$O. Numerous studies have since explored how planet formation and migration affect the C/O ratio of the planet \citep{madhusudhan_2014, mordasini_2016, cridland_2016, cridland_2017, ali-dib_2017, booth_2017, madhusudhan_2017}. Although there are differences in the composition of the model planets in these studies, there is general agreement that the partitioning of the carbon and oxygen between the gas and solid phases, along with the amount these species accreted, is important in determining the planet's composition. 

\smallskip

However, the composition of discs is not static in time, which needs to be taken into account in planet formation models. Although \citet{cridland_2016, cridland_2017} recently coupled a planet formation model to a disc model including chemical kinetics, they focused on planets that accreted their envelopes inside the water ice line, where the majority of volatile carbon and oxygen bearing species are in the gas phase. Further out in the disc, in regions where the temperatures are lower and molecular species freeze out, chemical kinetics can have a greater impact on the gas and solid phase composition by exchanging carbon and oxygen between species that are in ices on grains or in the gas phase. Using an extensive gas-grain network \citet{eistrup_2016} showed that CH$_4$ and CO can be removed from the gas phase, with the net effect of lowering the gas-phase C/O ratio inside of around 10~au. However, this process is slow, taking several million years \citep{eistrup_2018,bosman_2018b}, by which time gas giant must be well underway \citep{greaves_2010, najita_2014, manara_2018}.  

\smallskip

By comparison, it is becoming clear that the transport of molecular species is important on time-scales comparable to chemical processes. \citet{bosman_2018a} showed that even conservatively slow models of transport were enough to raise the CO$_2$ abundance above levels that has previously been assumed. Furthermore, transport is an essential component of a popular  new model for planet formation -- pebble accretion \citep{ormel_2010,lambrechts_2012} -- which relies on dust grains large enough that they decouple from the gas (e.g. \citealt{lambrechts_2014, morbidelli_2015, bitsch_2015}. \citealt{johansen_2018} invoke smaller pebbles, although they are still large enough to migrate). These pebbles thus migrate rapidly towards the star \citep{weidenschilling_1977} carrying the volatile molecular species frozen out onto their surfaces with them as they migrate. These volatile species then enter the gas phase when the pebbles cross their respective ice lines. Under the conditions often assumed in pebble accretion models, radial drift enhances the gas phase abundances by a factor of a few within a Myr \citep{booth_2017, krijt_2018}.

\smallskip

Previous studies investigating the interplay between chemical kinetics and transport typically assume that dust and gas are coupled (e.g. \citealt{aikawa_1999, tscharnuter_2007, nomura_2009, semenov_2011, heinzeller_2011, walsh_2014, gail_2017}; see \citealt{henning_2013}, for a review). These studies suggest that the transport has two predominant effects: 1) diffusion weakens concentration gradients and 2) transport limits the amount of time molecular species spend at a given location, reducing the influence of chemical processes that act on time scales longer than the transport time-scale. While a number of studies have treated the transport of gas and solids separately (e.g. in the context of CO sequestration in disc mid-planes; \citealt{piso_2015, piso_2016, bergin_2016, kama_2016, booth_2017, krijt_2018}; or the destruction of carbon grains in the terrestrial planet-forming region; \citealt{klarmann_2018}), to our knowledge this is the first study to treat gas and dust transport independently and gas-phase kinetics in detail. We emphasise that differences in the transport of gas and dust is the only way that the bulk atomic abundances (i.e. gas+dust abundance of carbon, oxygen, and nitrogen, etc) can vary in the disc; the partitioning of different molecular species between the gas and ice phase changes the atomic abundance of the each phase separately, but not the total abundance.

\smallskip

In this work we investigate the competition between chemical kinetics and transport in the mid-plane of protoplanetary discs. To this end, we couple a chemical kinetics network \citep{ilee_2011} with a 1D disc evolution model. Our disc evolution model includes gas evolution (which is treated as viscous), grain growth and radial drift \citep{booth_2017}\footnote{\url{https://github.com/rbooth200/DiscEvolution}}. We allow for differences in the transport of gas and ice phase molecular species, tracking their motion with the gas and dust, respectively. We examine how transport and chemical kinetics compete, focusing on the classical giant planet forming region (1--10~au). 

\section{Methods}

\subsection{Physical evolution}

We consider the physical evolution of gas and dust following \citet{booth_2017}. The gas evolution is treated assuming a viscous disc, with a constant $\alpha = 10^{-3}$ or $10^{-2}$. The justification of these choices is given in \autoref{sec:methods:phys_params}. Grain growth and radial drift are treated based upon \citet{birnstiel_2012}. The mid-plane temperature is computed taking into account viscous heating and irradiation from the central star, along with external irradiation with a temperature of $10\,{\rm K}$.

\smallskip

We self-consistently take into account the motion of each of the chemical species included in the model, treating the motion of gas and ice phase species independently. The gas phase species are advected at the gas velocity, which is determined by viscous evolution. Ice phase species are advected along with the dust. The dust radial velocity is set by a combination of both the gas velocity due to viscosity and radial drift \citep[e.g.][]{takeuchi_2002}.

The grain growth model assumes that there are two populations of grains, small and large, here we assume that the ice species are partitioned across the two populations in the same way as the dust. This results in the advection of ice phase species being dominated by the large grain population, in good agreement with calculations treating the full distribution of sizes \citep{stammler_2017}. In addition to advection, we include turbulent diffusion. The diffusion coefficient $D$ of the gas is given by $D = \nu / Sc$, where $\nu$ is the viscosity and the $Sc$ the Schmidt number, for which we  take $Sc = 1$. For the dust and ice phases the diffusion coefficient is scaled to the gas coefficient following \citet{youdin_2007} and assuming the eddy turnover time-scale equals the dynamical time-scale, $\Omega^{-1}$. For the dust sizes considered here, the diffusion coefficients of the dust and gas are the same to within one per cent.

\smallskip

The opacity used in the mid-plane temperature calculation has been updated. In \citet{booth_2017} we used the Rosseland mean opacity tables similar to \citet{bell_1994}, as computed by \citet{zhu_2012}. Since these opacities assume a grain size distribution appropriate for the interstellar medium they can considerably overestimate the dust opacity once the grains have grown and begin to migrate. Here we instead use Rosseland mean opacity computed self-consistently for grain size distributions with number density $n(a) \propto a^{-3.5}$ with the maximum grain size taken from the grain growth model. The dust properties follow \citet{tazzari_2016}, based on the composition of \citet{pollack_1994} and assuming a porosity of 30 per cent. 

\smallskip

In addition the models that include the effects of transport processes: viscous evolution, radial drift and diffusion; we also consider models that switch off some or all of these processes. This allows us to separate out the effects of physical and chemical processes on the composition.

\smallskip

\subsection{Physical parameters}
\label{sec:methods:phys_params}

\begin{figure*}
    \centering
    \includegraphics[width=\textwidth]{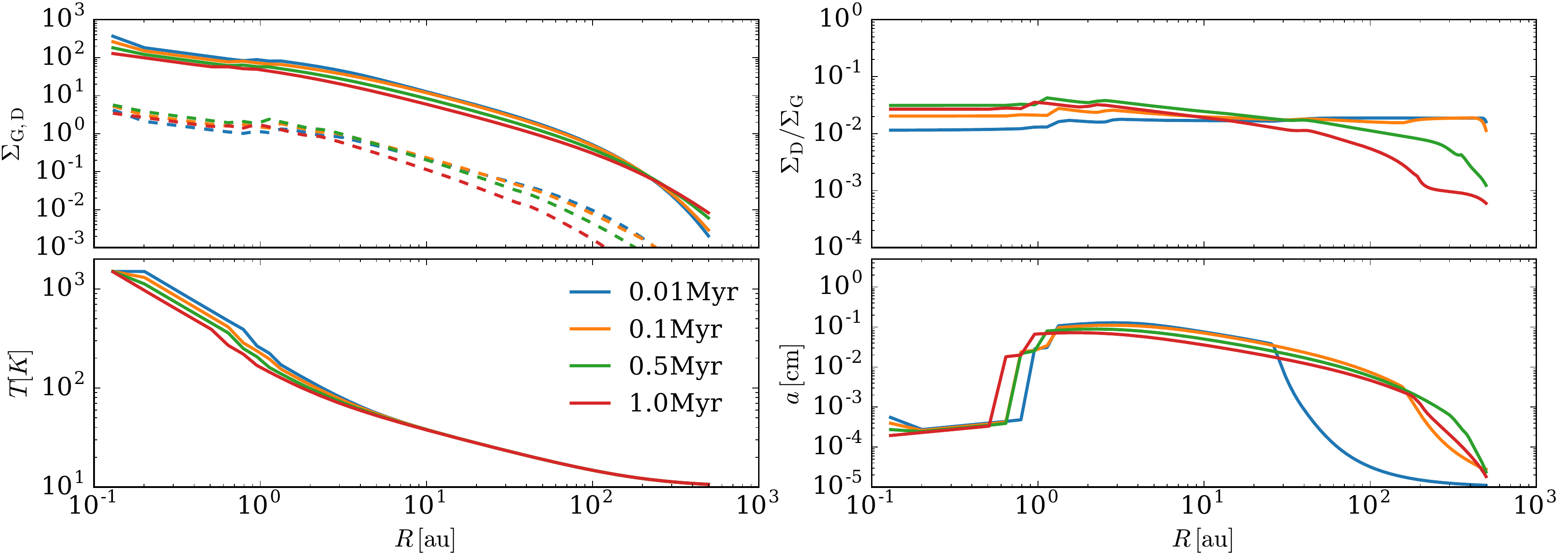}
    \caption{Evolution of the physical parameters for the model with $\alpha = 10^{-2}$. Top left: gas (solid) and dust (dashed) surface density. Top right: dust-to-gas-ratio. Bottom left: Temperature. Bottom Right: maximum grain size.}
    \label{fig:phys_evo_alpha2}
\end{figure*}
\begin{figure*}
    \centering
    \includegraphics[width=\textwidth]{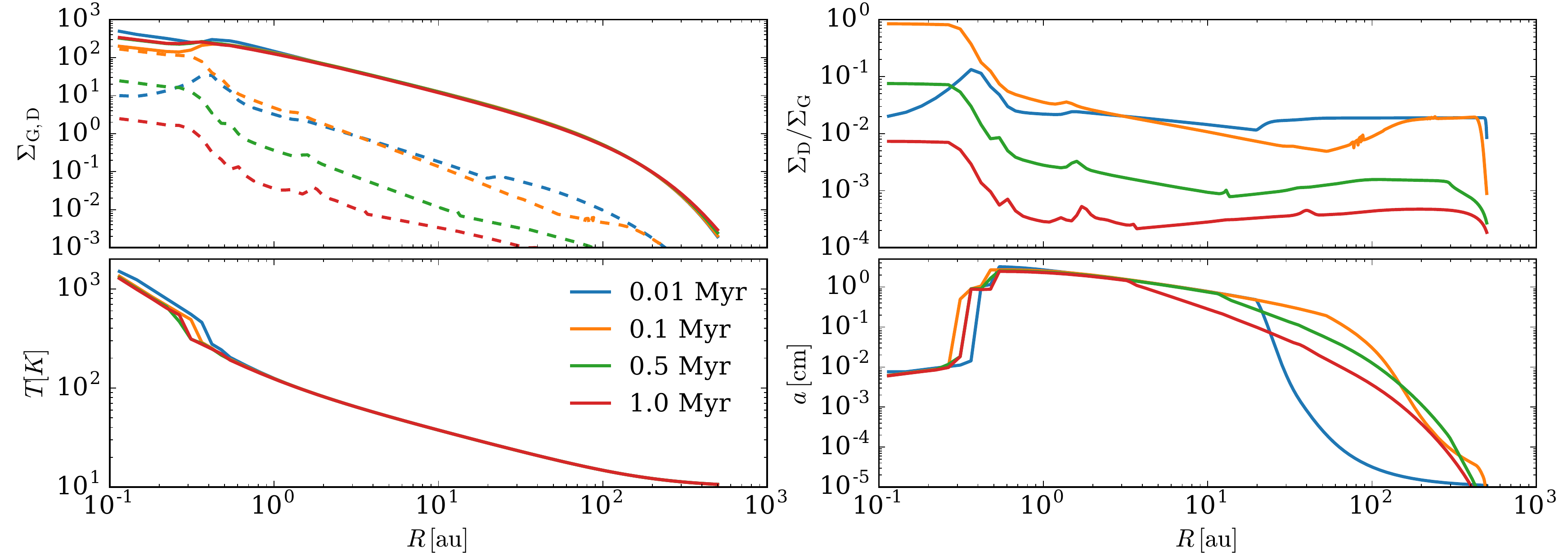}
    \caption{Same as \autoref{fig:phys_evo_alpha2}, but for $\alpha = 10^{-3}$.}
    \label{fig:phys_evo_alpha3}
\end{figure*}

We explore two models for the evolution of the disc, with the same initial surface density profile, but choices of the turbulent $\alpha$ parameter -- and correspondingly different accretion rates. Unless otherwise stated, in each model we set the initial disc mass to $0.01\,M_\odot$ and stellar mass to $1\,M_\odot$. The surface density profile is initialised to a \citet{1ynden-bell_1974} profile, 
\begin{equation}
\Sigma(R) = \frac{M_{\rm d}}{2 \upi R_d^2} \exp\left(-\frac{R}{R_{\rm d}} \right).
\end{equation}
We assume that the initial ratio of refractory dust-to-gas surface density is 0.01 everywhere. The total dust-to-gas ratio is higher than this by approximately factor of 2 due to the presence of ices frozen onto grain surfaces. The initial grain size is also taken to be 0.1\micron.

\smallskip

For the turbulent $\alpha$ parameter we use $\alpha = 10^{-3}$ and ${\alpha = 10^{-2}}$. While we note that recent observations suggest turbulence in the outer regions of discs is weaker than this \citep{pinte_2016,flaherty_2017, flaherty_2018, teague_2018}, there is evidence that the accretion rate through discs is more compatible with higher values of $\alpha$. In the absence of direct measurements of the radial flow of gas through discs, the accretion rate onto the star combined with the disc mass provide the most reasonable estimates. For example, \citet{boneberg_2016} found that an effective $\alpha \sim 0.1$ was required to match the accretion rate of HD 163296, while the turbulent mixing in the vertical direction was lower. Furthermore, based on the surface density profile and accretion rate \citet{clarke_2018} derive $\alpha \sim 0.01$ for CI Tau. Even TW Hya, which is an old system and has a low accretion rate, has an accretion rate compatible with $\alpha \sim 10^{-3}$ \citep{calvert_2002, bergin_2013,ercolano_2017}. Based on dust disc sizes and ages, \citet{andrews_2007} and \citet{guilloteau_2011} suggest $\alpha$ in the range $10^{-3}$--$10^{-2}$. Similarly, for most discs \citet{rafikov_2017} inferred $\alpha$ in the range $10^{-3}$--$10^{-2}$ based on the disc radii and masses inferred from dust emission in Lupus and their associated accretion rates. High dust-to-gas ratios in Lupus, as suggested by \citet{miotello_2017}, would result in higher estimates of $\alpha$ for Lupus; however, \citet{manara_2016} suggest that the dust mass is well matched to the accretion rate (but see \citealt{mulders_2017} and \citealt{lodato_2017} for a discussion). Differences between the accretion rate and inferred turbulence levels may point to MHD winds as the source of angular momentum transport instead of viscosity \citep{suzuki_2009, fromang_2013, bai_2013}. However, since we are mainly concerned with the speed of mass transport through the disc, this distinction is not critical. Thus we model the discs as viscous, considering $\alpha$ in the range $10^{-3}$--$10^{-2}$ to be typical for nearby, low mass protoplanetary discs.

\smallskip

For the dust evolution, we use the same parameters as \citet{booth_2017}, for which the standard parameters are based on fits to more detailed models \citep{birnstiel_2012}, except that we use the ice fraction to specify the location where the fragmentation velocity transitions from water ice-like ($10\,{\rm m\,s}^{-1}$) to silicate-like ($1\,{\rm m\,s}^{-1}$). Rather than varying these parameters, we choose instead to simply run models with radial drift included or neglected (in which case the dust moves with the gas). Combined with the two choices of $\alpha$, this spans the commonly studied parameter space. For example $\alpha = 10^{-3}$ is commonly considered \citep[e.g.][]{birnstiel_2012}, for which dust grows large enough to be in the radial drift dominated regime at large radii, leading to rapid evolution of the dust-to-gas ratio. This model is also comparable to the parameters assumed in pebble accretion models of \citet{lambrechts_2014, morbidelli_2015} and \citet{bitsch_2015}. In the $\alpha = 10^{-2}$ model, dust growth is initially limited by fragmentation everywhere, resulting in smaller grains and slower radial drift. Slower radial drift combined with a faster gas radial velocity results in the dust-to-gas ratio decreasing more slowly. Such a model may be in better agreement with observed disc properties, where low dust-to-gas ratios are typically not inferred (\citealt{boneberg_2016, ansdell_2016}; although note that masses determined from CO observations are uncertain), and is also comparable to the parameters assumed in the pebble accretion model of \citet{johansen_2018}.

\smallskip

The evolution of the surface density, dust-to-gas ratio, temperature, and maximum grain size for our two canonical models are shown in \autoref{fig:phys_evo_alpha2} ($\alpha = 10^{-2}$) and \autoref{fig:phys_evo_alpha3} ($\alpha = 10^{-3}$). The evolution of combined dust-gas models has been described at length in \citet{birnstiel_2012}, and \citet{booth_2017} in the case where adsorption and desorption at ice lines is included. Thus we only describe the salient differences here and refer the readers to the above references for further details. The main features of note are: the slower evolution of dust-to-gas ratio at $\alpha = 10^{-2}$, in which the dust-to-gas ratio is enhanced by a factor $\sim 2$ inside 10~au, and only decreases outside the scaling radius (100~au). In comparison the dust-to-gas ratio falls to $10^{-4}$ on Myr time-scales in the $\alpha = 10^{-3}$ model, a consequence of larger grain maximum grain size. Note also that gas surface density decreases in the $\alpha = 10^{-2}$ model, but barely changes after 1~Myr at $\alpha = 10^{-3}$. Another consequence of the higher accretion rate at $\alpha = 10^{-2}$ is the warmer temperature in the disc inside of $\sim 5\,{\rm au}$, due to both the larger contribution of viscous heating and also a higher optical depth because of smaller grains.

\smallskip

The time-scale for the evolution of the gas and dust surface densities are given by the time-scale for the accretion of gas and dust onto the star. These time-scales will be useful in the following sections for interpreting the competition between chemical evolution and transport. In the case of the gas, the accretion time-scale is controlled by the viscous velocity, $v_\nu$, from which we can estimate a viscous time-scale, $t_\nu = R / v_\nu$,
\begin{equation}
    t_\nu \approx 2\times 10^5 \left(\frac{\alpha}{10^{-3}} \right)^{-1} \left(\frac{R}{\rm au}\right)\,{\rm yr}.
\end{equation}
The long viscous time-scale at 100~au explains the slow change in gas surface density in $\alpha = 10^{-3}$ model. However, at 1~au the viscous transport time is always much less than 1~Myr, and for $\alpha = 10^{-2}$ the viscous transport time is less than 1~Myr everywhere inside of approximately 50 au.

In addition to transport with the gas, the dust surface density also evolves due to radial drift, with the associated radial drift time-scale, $t_{\rm drift} = R / v_{\rm drift}$. In this case, the evolution of the dust surface density is always on the shorter of radial drift and viscous time-scales. The radial drift time-scale depends on both the dust grain size and gas surface density profile. Taking the gas surface density to be $\Sigma = \Sigma_0 (R/{\rm au})^{-1}$, we estimate the radial drift time-scale as
\begin{equation}
t_{\rm drift} \approx 1 \times 10^5 \left(\frac{\Sigma_0}{10^2\,{\rm g\,cm}^{-2}}\right) \left(\frac{a}{\rm mm}\right)^{-1}\,{\rm yr}.
\end{equation}

Typical grain sizes are 0.1--1~mm for $\alpha = 10^{-2}$ and 1--10\,mm for $\alpha = 10^{-3}$. Hence the dust evolves on approximately the viscous time-scale for $\alpha = 10^{-2}$, which explains the slow evolution of dust-to-gas ratio. However, for $\alpha = 10^{-3}$ the dust evolves faster than the gas, and the dust-to-gas ratio decreases.

\subsection{Chemical evolution}

Within the physical model discussed above, we embed a time-dependent gas-grain chemical evolution model that self-consistently calculates the chemical evolution of each grid cell.  For this, we utilise the {\sc krome} chemical evolution package\footnote{\url{http://www.kromepackage.org/}} \citep{grassi_2014}.  For the chemical reaction network, we adopt the network previously used in \citealt{ilee_2011, evans_2015} and \citealt{ilee_2017}, with modifications described here to provide consistency with more recent disc evolution models. {The network consists of 1485 reactions involving 136 species} containing the elements H, He, C, N, O, Na and S.  These reactions were originally selected from a subset of the UMIST Rate 95 database \citep{millar_umist_1997}. Data from the Kinetic Database For Astrochemistry, KIDA\footnote{\url{http://kida.obs.u-bordeaux1.fr}}, were used to update some of the rates and rate coefficients.  In addition to the chemical reactions, adsorption, and desorption processes; we self-consistently follow the advection and diffusion of each of the gas and ice phase species following \citet{booth_2017}.  We overview the key aspects of our chemical model below.   

\smallskip

The fundamental variables integrated in the disc evolution and transport code are the total surface density, dust fraction, and mass fraction of each chemical species. These are used to compute the mid-plane number density, $n(i)$, and dust-to-gas ratio, $\epsilon$, assuming a Gaussian vertical structure using the scale-height computed from the mid-plane temperature. The fractional abundance relative to the total number of hydrogen nuclei, $n$, is $X(i) \equiv n(i) / n$.  The rate equation for the gas-phase fractional abundance of species $i$ is therefore
\begin{eqnarray}
\frac{\rm{d}}{\rm{d}t} X(i) & \equiv & S(i) \label{eqn:RATES} \\
\nonumber &=& \sum_{j,l,m} k(j) X(l) X(m) n \\
\nonumber &-& \sum_{j',m} k(j') X(i) X(m) n \\
\nonumber &-& 2 \sum_{j''} k(j'') X(i)^{2} n \\ 
\nonumber &-& S_{\rm{rad}}(i)X(i) + S_{3}(i) + S_{\rm d}(i)- S_{\rm a}(i),
\end{eqnarray}
where $k(j)$ is the rate coefficient of the $j$th reaction, and the summations are restricted so that only reactions involved in the formation or removal of the $i$th species are included. We denote this total rate $S(i)$. These rate coefficients are of the standard Arrhenius form
\begin{equation}
k(j)=\alpha(j) \, \left(\frac{T}{300}\right)^{\beta(j)} \, \exp{\left(\frac{-\gamma(j)}{T}\right)},
\end{equation}
where $\alpha(j)$ is the room temperature rate coefficient of the reaction (at $300\,$K), $\beta(j)$ describes the temperature dependence and $\gamma(j)$ is the activation energy of the reaction. 
The terms $S_{\rm{rad}}(i)X(i)$, $S_{3}$, $S_{\rm a}$ and $S_{\rm d}$ represent radiative destruction processes, three-body gas phase reactions, adsorption on to dust grains, and desorption from dust grains, respectively.  

\smallskip

For three body reactions, we assume that the only third body of importance is H$_2$, and therefore
\begin{eqnarray}
S_3(i) &=& \sum_{j,l,m} k(j) X(l) X(m) X({\rm{H}_{2}}) n^{2} \\
\nonumber &-& \sum_{j',m} k(j') X(i) X(m) X({\rm{H}_{2}}) n^{2} \\
\nonumber &-& 2 \sum_{j''} k(j'') X(i)^{2} X({\rm{H}_{2}}) n^{2}.
\end{eqnarray}

\smallskip

The rate of destruction of species $i$ due to radiative processes is given by
\begin{equation}
S_{\rm rad} = \alpha(i) \, \exp(-\gamma A_{\rm V})
\end{equation}
for photoreactions (where $\alpha$ is a proportionality constant containing the dust grain albedo, which we take to be 0.5).  For cosmic ray ionisation, we take $S_{\rm rad} = \zeta = 10^{-17}\,\rm{s}^{-1}$ everywhere, unless otherwise specified.  Though photoabsorption of radiation from external sources will affect the chemistry in the upper layers of a disc, we focus on the bulk of the planet forming material toward the disc mid-plane. We therefore assume that all material is well shielded from external sources of photons, setting $A_{\rm V} = 30$\,mag everywhere. However, we do include photoreactions due to secondary photons generated by cosmic rays, using the rates from \citet{heays_2017}.

\smallskip

For the adsorption of species from the gas phase onto the surfaces of icy grain mantles, we assume
\begin{equation}
S_{\rm{a}} = X_{\rm dust} \, S(i) \, \sqrt{\frac{8kT}{\pi m(i)}} \, \langle \pi a^{2} \rangle   
\label{eqn:ads}
\end{equation}
where $X_{\rm dust}$ is the dust number density, $S(i)$ is the sticking coefficient (taken to be 0.3), $m(i)$ is the atomic mass of the adsorbed species, and $\langle \pi a^{2} \rangle   $ is the average dust grain area. The average grain radius is computed using the maximum grain size from the dust evolution code, assuming a grain size distribution $n(a) {\rm d}a \propto a^{-3.5} {\rm d} a$ with the minimum grain-size equal to $0.1\micron$. However, we note that our results are only very weakly dependent on this choice (as shown in \autoref{App:Params}).

\smallskip

For unsaturated C, N and O species on the surfaces of dust grains, we assume hydrogenation occurs via the Eley-Rideal mechanism, at the rate of adsorption of H from the gas multiplied by the probability of encounter with the species on the grain surface, e.g., gC $\rightarrow$ gCH $\rightarrow$ \ldots $\rightarrow$ gCH$_4$ (where g denotes a species on a grain surface, see \citealt{visser_2011}).

\smallskip

For the thermal desorption of species from dust grain surfaces into the gas phase, we assume
\begin{equation}
S_d(i) = 1.26\times10^{-21} \, \sigma \, \nu_{0}(i)
\; \exp \left( \frac{-E_{\rm{b}}(i)}{k T} \right),
\label{eqn:des}
\end{equation}
where $\sigma = 1.5\times10^{15}\,$cm$^{-2}$ is the surface density of binding sites, $E_{\rm{b}}(i)$ is the binding energy of the $i$th species on the surface of the dust grain, and $T$ is the dust temperature (which we assume to be in equilibrium with the gas temperature), and $\nu_{0}$ is the characteristic vibrational frequency of the species attached to the grain.

\smallskip

Integration of the rate equations in each cell is performed independently, using the \textsc{dlsodes} package \citep{hindmarsh_1983} within the \textsc{krome} package \citep{grassi_2014} to yield the time-dependent evolution of fractional abundance for each species throughout the disc. 

\subsection{Combined chemical evolution and transport}

The coupling to the disc evolution model is achieved in a a first-order operator-split fashion, which we briefly describe here. The full equation for the surface density of a given species, $\Sigma(i)$, may be written as
\begin{equation}
    \pderiv{\Sigma(i)}{t} + \frac{1}{R}\pderiv{}{R}\left[R \Sigma(i) v_r(i) \right] = \Sigma(i) S(i),
\end{equation}
where $v_r(i)$ is the radial velocity of the species (either the gas velocity or dust velocity, depending on whether the species is in the gas or ice phase), and $S(i)$ is the source terms due to chemical reactions (\autoref{eqn:RATES}). We discretize this equation on a fixed Eulerian grid, as described in \citet{booth_2017}. We then proceed by splitting these equations into two steps, first a transport step without the chemical reactions:
\begin{equation}
    \pderiv{\Sigma(i)}{t} + \frac{1}{R}\pderiv{}{R}\left[R \Sigma(i) v_r(i) \right] = 0,
\end{equation}
where the amount of each species being transported between the different radial cells between times $t$ and $t+{\rm d}t$ is computed as described in \citet{booth_2017}. This produces a new estimate for the surface density in each cell, $\Sigma^*(i)$, at time $t+{\rm dt}$. 

\smallskip

The second step in the computation is to integrate 
\begin{equation}
    \pderiv{\Sigma(i)}{t} = \Sigma(i) S(i)
\end{equation}
over a time ${\rm d}t$ starting from $\Sigma^*(i)$. To do this, we first compute mid-plane density of each chemical species via $\rho(i) = \Sigma^*(i) / (\sqrt{2\upi} H)$,  where $H$ is the disc-scale height. We then pass $n(i) = \rho(i) / m(i)$, where $m(i)$ is the mass of the species, into the chemical solver, \textsc{krome}, which integrates the number density from $t$ to $t + {\rm d} t$. The new $n(i)$ can then be converted back to $\Sigma(i)$.  Finally, the grain size is updated as described in \citet{booth_2017} and the temperature in the disc updated.

\subsection{Chemical initial conditions}
\label{sec:initial}

For the initial abundances, we assume the gas is molecular with the abundance of key molecular species given in \autoref{tab:init_abund}, which follow \citet{eistrup_2016}, which are comparable to the abundances in comets \citep{mumma_2011}.  These abundances correspond to C/H, N/H, and O/H ratios of 0.47, 0.89, and 0.85 times solar, respectively \citep{asplund_2009}. The C/O ratio is  0.29, which is below the protosolar value of approximately 0.54. These lower abundances are consistent with the remaining species being locked up in refractory solids \citep{pollack_1994}. Where relevant, we thus assume the remaining C, N, and O atoms are locked up in the cores of dust grains, which for the purpose of this study are considered inert.

\begin{table}
    \centering
    \caption{Initial abundances relative to the number of hydrogen nuclei, $X(i) \equiv n(i) / n$. Based on \citet{eistrup_2016}.}
    \begin{tabular}{lc | lc}
        \hline
        Species & $X(i)$ &Species & $X(i)$ \\
        \hline
        \hline 
         H       & $9.11\times 10^{-5}$ & CH$_4$  & $1.8\times 10^{-5}$    \\
         H$_2$   & $0.4999545$          & N$_2$   & $2.1\times 10^{-5}$   \\
         He      & $0.098$              & NH$_3$  & $2.1\times 10^{-5}$   \\
         H$_2$O  & $3.0\times 10^{-4}$  & H$_2$S  & $6.0\times 10^{-6}$    \\
         CO      & $6.0\times 10^{-5}$  & Na      & $3.5\times 10^{-5}$    \\
         CO$_2$  & $6.0\times 10^{-5}$  & & \\
         \hline
    \end{tabular}
    \label{tab:init_abund}
\end{table}

\section{Chemical evolution}
\label{sec:results:chem}

\subsection{Chemical evolution without transport}
\label{sec:results:chem:static}

\begin{figure*}
    \centering
    \includegraphics[width=\textwidth]{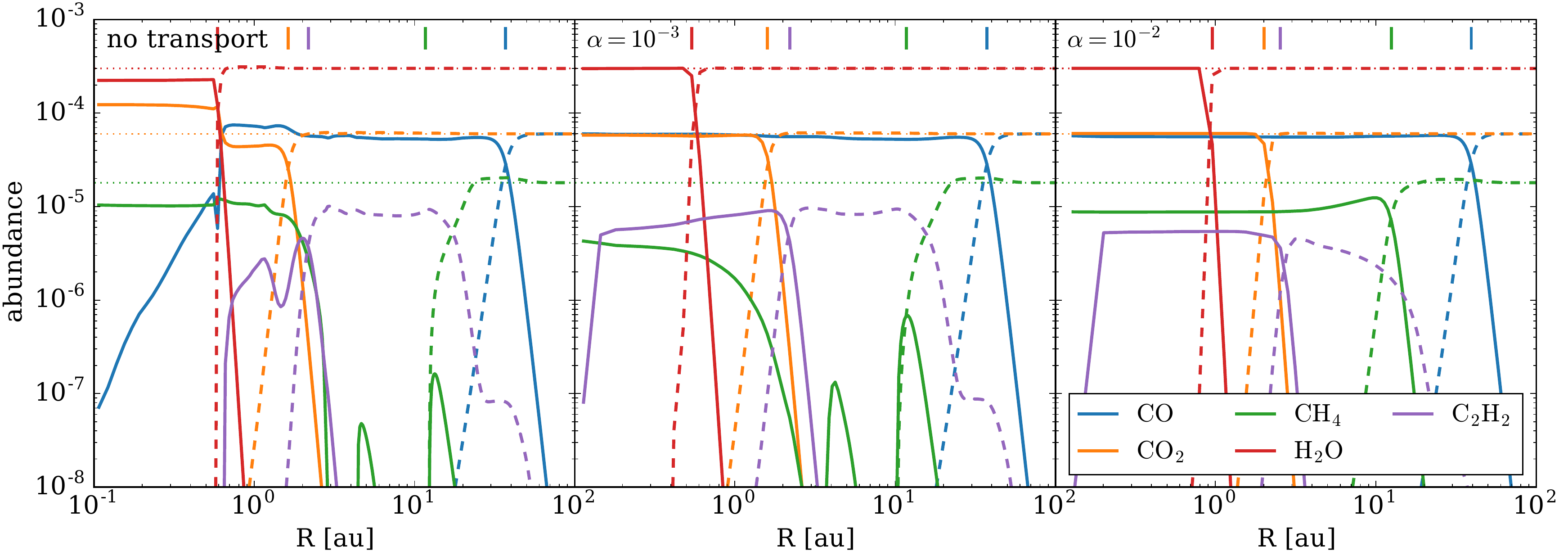}
    \caption{Abundances relative to hydrogen, $X(i)$, of the major carbon and oxygen carriers after $10^6\,{\rm yr}$ in models without radial drift. In the left panel transport is neglected entirely, while the other two panels show models in which the dust moves at the same speed as the gas. Solid lines denote gas phase species and dashed lines denote ice phase species and the dotted lines show the initial abundance of each species. The vertical bars show the approximate ice line locations. Note that C$_2$H$_2$ should be considered a proxy for hydrocarbons more generally.}
    \label{fig:static_chem}
\end{figure*}

We first begin with a simple test case in which all transport terms are turned off in the disc evolution model, i.e.\ the chemical reaction network for each cell is integrated independently with the density fixed at the initial value. This allows a direct comparison with recent works exploring the impact of chemistry on the composition of planets \citep{eistrup_2016,cridland_2016,yu_2016,cridland_2017,eistrup_2018}, with which we can benchmark the smaller chemical network used in this study. For this test we present results from the standard disc model with a mass of ${0.01 M_\odot}$ and $\alpha = 10^{-3}$. The evolution in these models is almost entirely driven by ionization due to cosmic rays (see Appendix \ref{App:Params} or \citealt{eistrup_2016}).  The abundances of the dominant carbon and oxygen carriers at 0 and $10^6\,{\rm yr}$ are shown in \autoref{fig:static_chem}, which can be compared with Figures 2 and 3 of \citet{eistrup_2016}. The time evolution can also be seen in \autoref{fig:ion_evo}.

\smallskip

Our model without transport is in good qualitative agreement with the results from the more extensive networks used by \citet{eistrup_2016, eistrup_2018}, and \citet{yu_2016}. The most significant effect is the depletion of CH$_4$ inside of 10~au, where CH$_4$ is in the gas phase. The primary pathway for the removal of CH$_4$ are reactions with C$^+$, which in turn is produced via reactions between CO and He$^+$ that is generated by cosmic ray ionization. The reactions between CH$_4$ and C$^+$ produce C$_2$H$_2^+$ and C$_2$H$_3^+$, which can recombine with electrons to form C$_2$H and C$_2$H$_2$ \citep{walsh_2015, yu_2016}. C$_2$H$_2$ then freezes out, acting as the main reservoir in our network. In reality, C$_2$H$_2$ is likely further hydrogenated on the grain surfaces -- \citet{eistrup_2018} find that the main reservoir in this region is the saturated hydrocarbon C$_2$H$_6$, which is also frozen out. We therefore argue that the conversion of CH$_4$ to larger molecules is adequately captured in our model because the time-scale for depletion (a few 1$0^5\,{\rm yr}$) is captured properly. Furthermore, since the binding energies of these two species are similar (\citealt{collings_2004, oberg_2009}, see also \citealt{penteado_2017}) they will be in the ice phase for similar ranges of parameter space, resulting in similar transport efficiency too. We thus argue that the differences are not significant for our goal, i.e. to asses how the carbon and oxygen abundance of the gas and solids during planet-formation is affected by the competition between transport and chemistry.

\smallskip

The most significant differences come from the fact that we do not include detailed grain surface chemistry. However, besides the aforementioned hydrogenation of small hydrocarbons, these differences are most significant on time-scales of a few Myr. E.g. \citet{eistrup_2018} show that this leads to the conversion of CO into CO$_2$ on time-scales of a few Myr  in the outer disc where CO is frozen out \citep[see also][]{bosman_2018b}. Without grain surface chemistry CO can only be destroyed in the gas phase, which takes several to $10\,{\rm Myr}$. For this reason we focus on the first $1\,{\rm Myr}$, where the differences in the CO abundance are within a factor of 2. 

\begin{figure*}
    \centering
    \includegraphics[width=\textwidth]{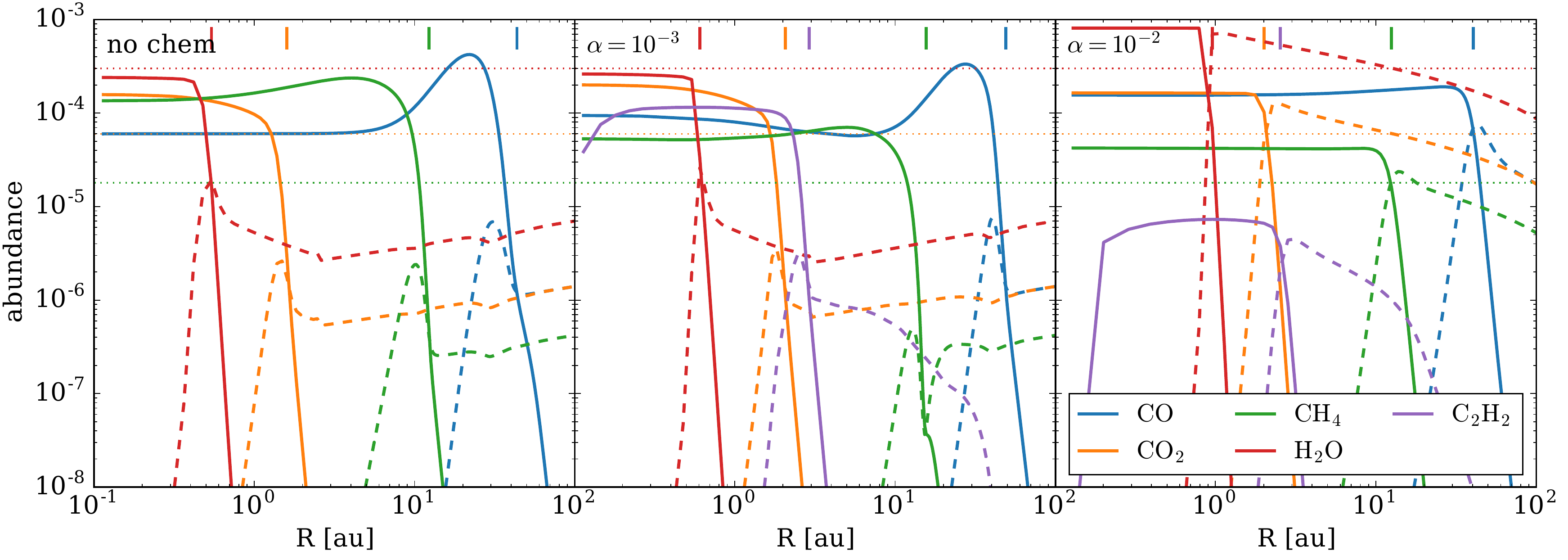}
    \caption{Same as \autoref{fig:static_chem}, but for models including radial drift. The left hand panel shows the $\alpha = 10^{-3}$ case in including only adsorption and desorption.}
    \label{fig:drift_chem}
\end{figure*}

\subsection{Viscously evolving discs}

We now turn our attention to models involving transport, first considering models with viscous evolution but no radial drift. In this model, the dust and gas move together as they move towards the star, similar to the models by \citet{aikawa_1999}. The differences in the chemical evolution between the models with and without transport are thus due to the gas at a given location having come from larger radii in the disc, where the temperatures and densities are lower. Since each parcel of gas now spends a limited amount of time in the regions where a given chemical process is happening (e.g. the depletion of CH$_4$), this naturally leads to competition between the chemical time-scales and transport time-scales (which are given in \autoref{sec:methods:phys_params}).

\smallskip

Transport thus reduces the impact of chemical reactions on the composition in the inner regions. Even in our low accretion rate model with $\alpha = 10^{-3}$, viscous transport carries gas from 2~au onto the star more rapidly than CO$_2$ depletes ($\gtrsim 10^6\,{\rm yr}$), as discussed in detail by \citet{bosman_2018a}. However, at $\alpha=10^{-3}$, viscous transport is not able to overcome the depletion of CH$_4$, due to a combination of a shorter chemical time-scale (a few $10^5\,{\rm yr}$) and the longer time that material spends at 10~au than at 2~au. 

At $\alpha = 10^{-2}$, viscous transport becomes efficient enough that gas-phase chemistry no longer has a significant effect on the abundance of the major carbon and oxygen carriers inside 10~au. While there is still some conversion of CH$_4$ to larger hydrocarbons, this only reduces the abundance of CH$_4$ by a factor of $\sim 2$. Thus the composition of the mid-plane inside the CO ice line ($\sim 30\,{\rm au}$) is largely set by the composition of the material being carried in from larger radii.

\subsection{Influence of radial drift}

The primary effect of radial drift is to enhance the transport of molecular ices from the outer disc \citep{booth_2017}. The molecular species frozen out enter the gas phase when the grains pass each species' respective ice lines, enhancing the local volatile abundance inside the ice line. The strength of this enhancement is controlled by the relative time-scales on which the gas and dust evolve: rapid radial drift leads to a rapid increase of volatile abundance inside the ice line. In the opposite limit, when radial drift is slow, there is no enhancement of the abundances because the gas and dust move together. Thus the volatile species are transported away from the ice lines in the gas phase at the same rate they are brought there in ices on the grains.

\smallskip

The influence of radial drift can be most clearly seen in the left hand panel of \autoref{fig:drift_chem}, which shows the $\alpha = 10^{-3}$ model including radial drift and adsorption/desorption, but neglecting other chemical processes. The rapid dust transport in this model means that 90 per cent of the volatiles are delivered into the inner disc within 0.5~Myr. Combined with the low gas accretion rate, this results in high gas phase molecular abundances in the inner disc, along with a reduced dust-to-gas ratio (\autoref{fig:phys_evo_alpha3}). 

\smallskip

\begin{figure*}
    \centering
    \includegraphics[width=\textwidth]{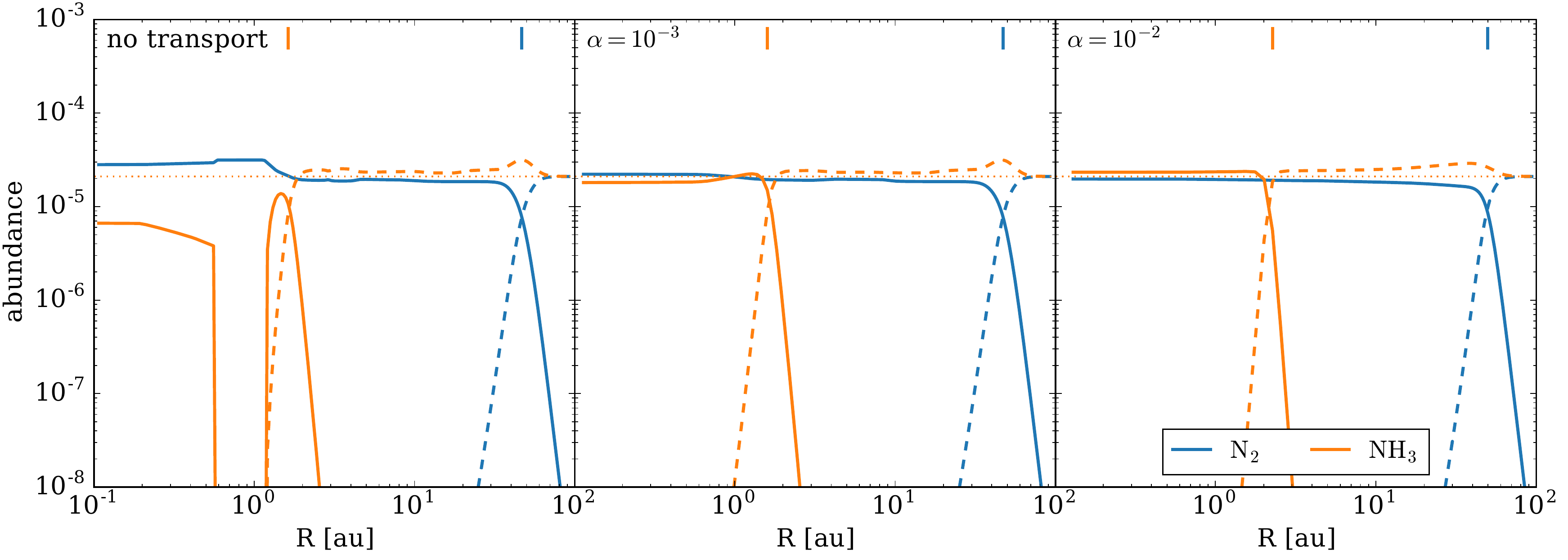} \\
    \includegraphics[width=\textwidth]{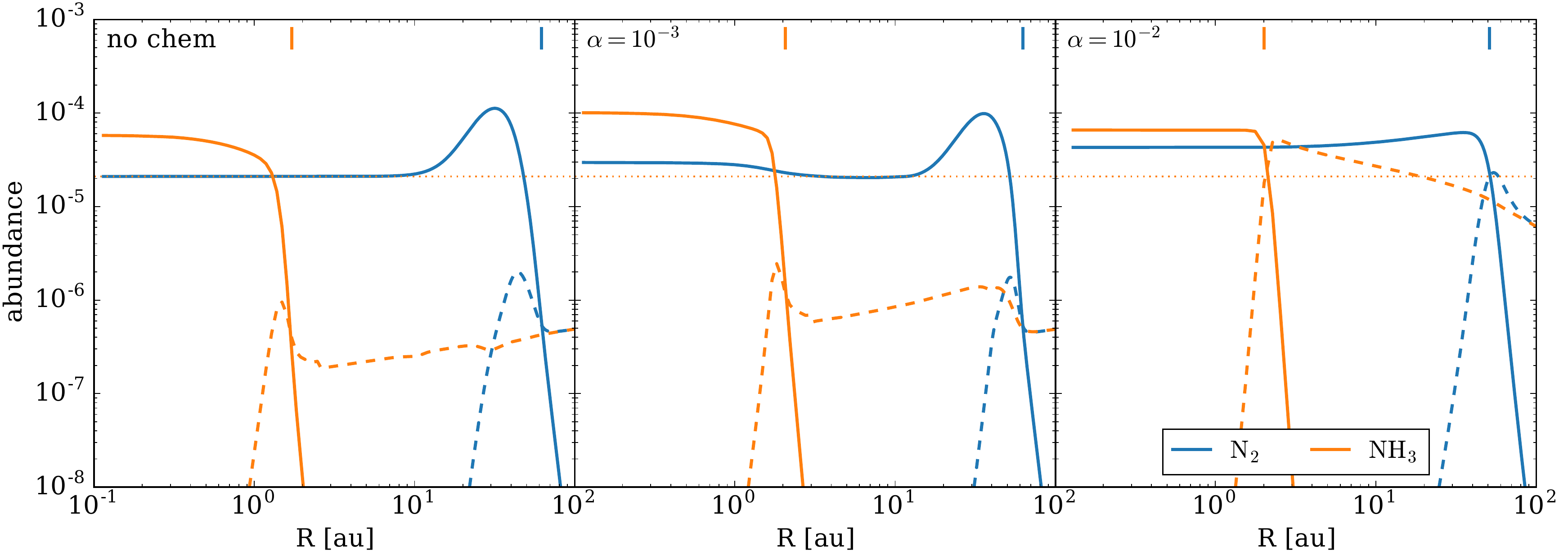}
    \caption{Abundances relative to hydrogen of the key nitrogen bearing molecules. The models are the same as \autoref{fig:static_chem} and \autoref{fig:drift_chem}. Solid lines denote gas phase species, dashed lines denote ice phase species. Top: models with viscous transport only. Bottom: models including radial drift.}
    \label{fig:N_chem}
\end{figure*}

For CO, which enters the gas phase at around 40~au where viscous transport is slow, this results in a spike in the gas phase CO abundance. Inside 5~au the CO abundance is dominated by gas that originated inside the ice line and has been carried inwards by the gas; thus, the CO abundance returns to its initial value. For the species that have their ice lines closer to the star, such as CH$_4$, CO$_2$ and H$_2$O, the abundance inside the ice lines has become close to constant because the transport time-scale is less than $1\,{\rm Myr}$. Outside their ice lines, the gas phase abundance of all molecular species remains low because any molecules that diffuse beyond the ice line freeze out and are carried back on the grain surfaces. This results in the ice-to-dust abundance of individual species being enhanced by as much as a factor of $\sim 10$ near their ice lines.

\smallskip

The interplay between gas-phase chemistry, radial drift and viscous evolution can be seen in the middle panel of \autoref{fig:drift_chem}, for the model with $\alpha = 10^{-3}$. Here we see again that the conversion of CH$_4$ to larger hydrocarbons is efficient inside of 10~au, removing the spike in CH$_4$ at the ice line. However, radial drift is efficient enough that the the CH$_4$ abundance is still high after 1~Myr. The high CH$_4$ abundance at $\sim 10\,{\rm au}$ also translates into a higher rate of hydrocarbon formation, which can be seen through the higher abundance of C$_2$H$_2$ ice relative to water ice in the middle panel of \autoref{fig:drift_chem} compared to \autoref{fig:static_chem}. However, efficient radial drift prevents the newly formed ice reaching high abundances, with the ice instead being transported in to its own ice line, as suggested in \citet{booth_2017}. A similar process is responsible for the higher CO abundance in the inner region. In this case it is the formation of H$_2$CO from CO near the ice line that is responsible. Once formed, the H$_2$CO freezes out and is transported on the grains to around 5~au, where it enters the gas phase and is converted back to CO. The enhancement of CO in the inner regions thus relies on the formation of H$_2$CO. We note that when grain surface reactions are included  \citet{eistrup_2016} find CO$_2$ is formed instead. While CO$_2$ would freeze out also and be transported inwards, the conversion of CO$_2$ into CO is slow compared to the transport. Thus the effect of CO conversion to CO$_2$ and transport on the grain surfaces would be to further enhance the CO$_2$ abundance in the inner regions, rather than CO. 

\smallskip

The evolution of species such as CO, CH$_4$, CO$_2$ and H$_2$O after $1\,{\rm Myr}$ would be characterized with an inside-out depletion of the species in order of their snow line locations, because species closest to the star accreted onto the star first \citep{booth_2017}. Including chemical reactions only slightly modifies this picture, with the depletion of CO and CH$_4$ being slightly accelerated with respect to the results without chemical reactions through their processing to CO$_2$ and larger hydrocarbon ices, respectively.

\smallskip

In the $\alpha = 10^{-2}$ model the evolution is similar both with and without radial drift. This is due to the viscous time-scale ($t_\nu$) and radial drift time-scale ($t_{\rm drift}$) being comparable, so radial drift only modestly affects the transport of dust. Furthermore, both of the transport time-scales are shorter than the chemical time-scales for the depletion of the major carbon and oxygen carriers. At $\alpha = 10^{-2}$ the impact of  radial drift is to increase the abundances by a factor $\sim 3$ due to the radial drift of ices. The enhancement in molecular abundances is only slightly higher than the enhancement in dust-to-gas ratio because the dust drifts inwards at only slightly higher speeds than the gas. However, suppressing the influence of radial drift entirely would require $t_\nu \ll t_{\rm drift}$. Similar to the $\alpha = 10^{-2}$ models without radial drift, we again see that gas-phase reactions only have a modest effect on the abundances, although the abundance of hydrocarbons produced from CH$_4$ still reaches 20 per cent of CH$_4$ abundance. 

\smallskip

In \autoref{fig:N_chem} we show how transport influences the evolution of the main nitrogen reservoirs, N$_2$ and NH$_3$. In the outer disc we find that the evolution is slow(time-scales $\gg1\,{\rm Myr}$), even in the absence of transport. We find that N$_2$ converts to NH$_3$, as found by \citet{schwarz_2014} in models with a high initial N$_2$ abundance. Conversely, \citet{eistrup_2016} report the conversion of NH$_3$ to N$_2$ in the outer disc, again on time-scales much longer than $1\,{\rm Myr}$. The difference is due to the destruction of NH$_3$ ice by cosmic-ray induced photons, which is neglected in this work. Nevertheless, the evolution of the nitrogen abundance in the outer disc is expected to be slow.

In the inner disc NH$_3$ is depleted and the N$_2$ abundance increases. The remaining NH$_3$ abundance inside the NH$_3$ ice line is sensitive to the rate of NH$_3$ production from N$_2$. We find that this is fastest via the pathway
\begin{equation}
{\rm N}_2\overset{\gamma_{\rm CR}}{\rightarrow} {\rm N}^+ \overset{{\rm H}_2}{\rightarrow}  {\rm NH}^+ \overset{{\rm H}_2}{\rightarrow} {{\rm NH}_2^+} \overset{{\rm H}_2}{\rightarrow} {{\rm NH}_3^+} \overset{{\rm H_2O}}{\rightarrow}  {{\rm NH}_4^+} \overset{{\rm e}^-}{\rightarrow} {{\rm NH}_3}.
\end{equation}
Outside the water ice line where water is not present in the gas phase the ${{\rm NH}_3^+} \rightarrow  {{\rm NH}_4^+}$ step occurs more slowly via reactions with H$_2$, resulting in lower abundances. In the colder regions further out NH$_3$ formation can also occur on grain surfaces. 

\smallskip

As in the case of carbon and oxygen bearing species, we see that even low levels of transport prevent the depletion of NH$_3$, largely because the depletion time-scale, $\sim 0.5\,{Myr}$, is long compared to the transport time-scales in the inner disc. We note that high NH$_3$ abundances inside the NH$_3$ ice line is in conflict with recent measurements of the NH$_3$ abundance in the inner regions of discs ($< 10^{-7}$; \citealt{pontoppidan_2019}), even for $\alpha = 10^{-3}$. However, it may be possible that the NH$_3$ abundance observed in the upper layers is not representative of the bulk abundance on the disc. A similar conflict between observed CO$_2$ abundances and models including transport was reported by \citet{bosman_2018a}.

\smallskip

Neither the conversion of N$_2$ to NH$_3$ nor the reverse process change the total gas phase nitrogen abundance significantly -- in the inner disc both species are in the gas phase, while in the outer disc the destruction of N$_2$ is slow. However, due to the long viscous time in the outer disc (1~Myr at 50~au for $\alpha=10^{-2}$), transport only has a modest effect on the evolution of nitrogen in the outer disc. Conversely, in the inner disc where the viscous time is much shorter, transport reduces the conversion of NH$_3$ to N$_2$. 

\smallskip

Models with radial drift also have a similar effect on the nitrogen reservoirs as the carbon and oxygen reservoirs. Radial drift causes enhancements in the gas phase N$_2$ abundance inside the N$_2$ ice line by a factor 2 -- 3. The NH$_3$ abundance increases by factors of 3--10 due to radial drift and is further enhanced by the formation of extra NH$_3$ in the outer disc, which freezes out and is carried in by the drifting grains. 

\smallskip

We briefly consider how sensitive the results presented here are to assumptions about the disc model. Notably, the radial drift and viscous time-scales are not very sensitive to model assumptions, apart from the dependence on $\alpha$. For this reason, we investigated how the chemical time-scales vary in models without transport, varying the disc mass, cosmic-ray ionization rate and average grain size used in the chemical models. The results are presented in detail in Appendix \ref{App:Params}, which shows that the time-scale for the depletion of CH$_4$ and CO are only sensitive to the cosmic-ray ionization rate. Since
we adopt the canonical value of cosmic-ray ionization rate ($10^{-17}\,{\rm s}^{-1}$), which assumes no shielding of the interstellar cosmic ray flux via stellar winds or magnetic fields \citep{cleeves_2013}, our results likely represent a lower limit of the importance of transport relative to mid-plane chemical kinetics.

\subsection{Abundance ratios}

\begin{figure*}
    \centering
    \includegraphics[width=\textwidth]{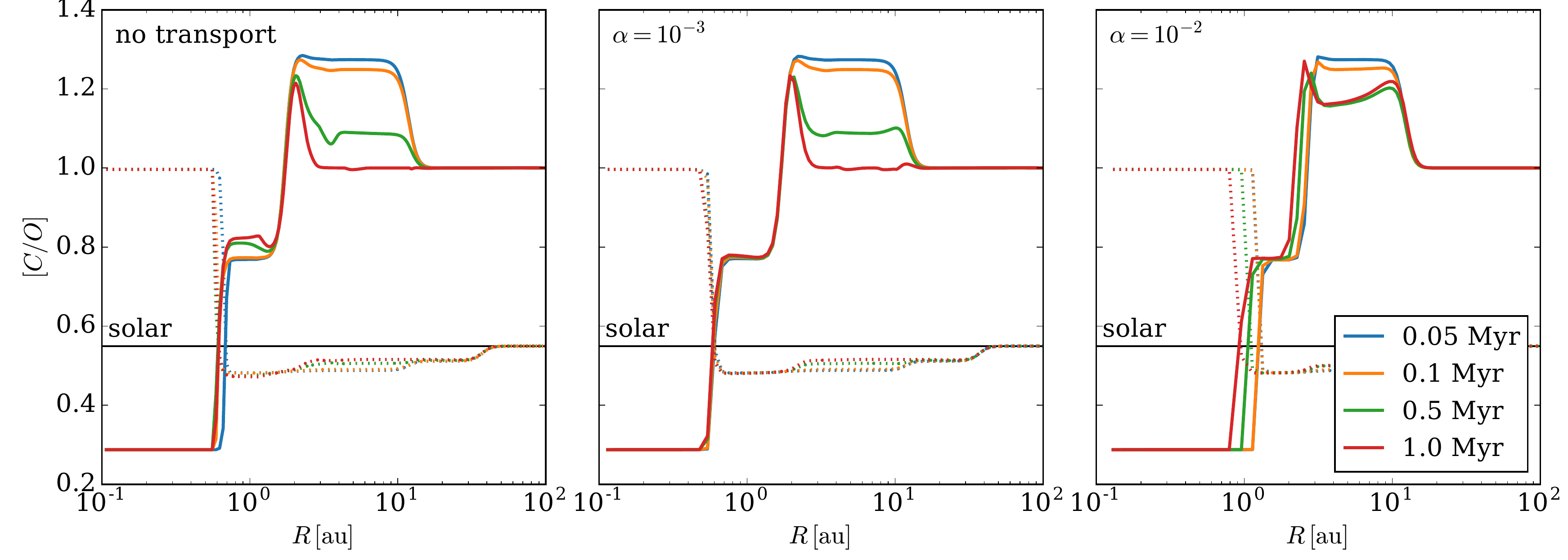}
    \caption{Evolution of the C/O ratio for models with viscous transport, but without radial drift or diffusion. For comparison the left hand panel shows the evolution without transport. The solid lines denote the gas phase abundances, while the dotted lines show the ice abundances.}
    \label{fig:visc_CO}
\end{figure*}

\begin{figure*}
    \centering
    \includegraphics[width=\textwidth]{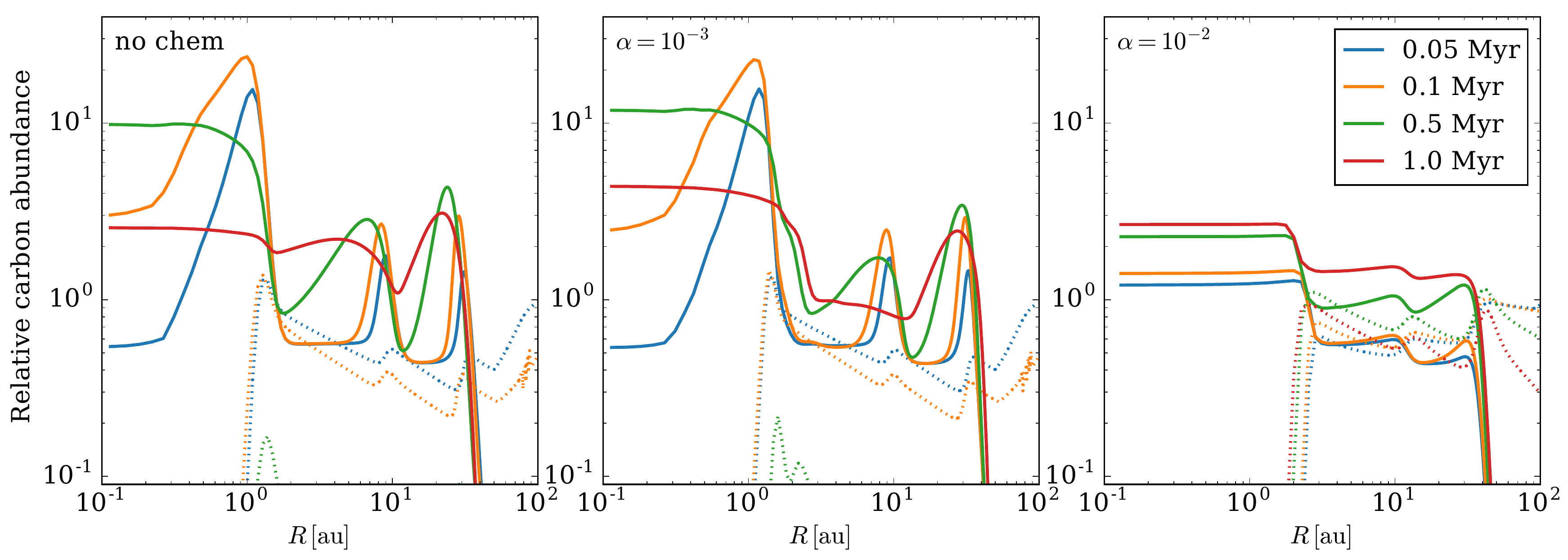} \\
    \includegraphics[width=\textwidth]{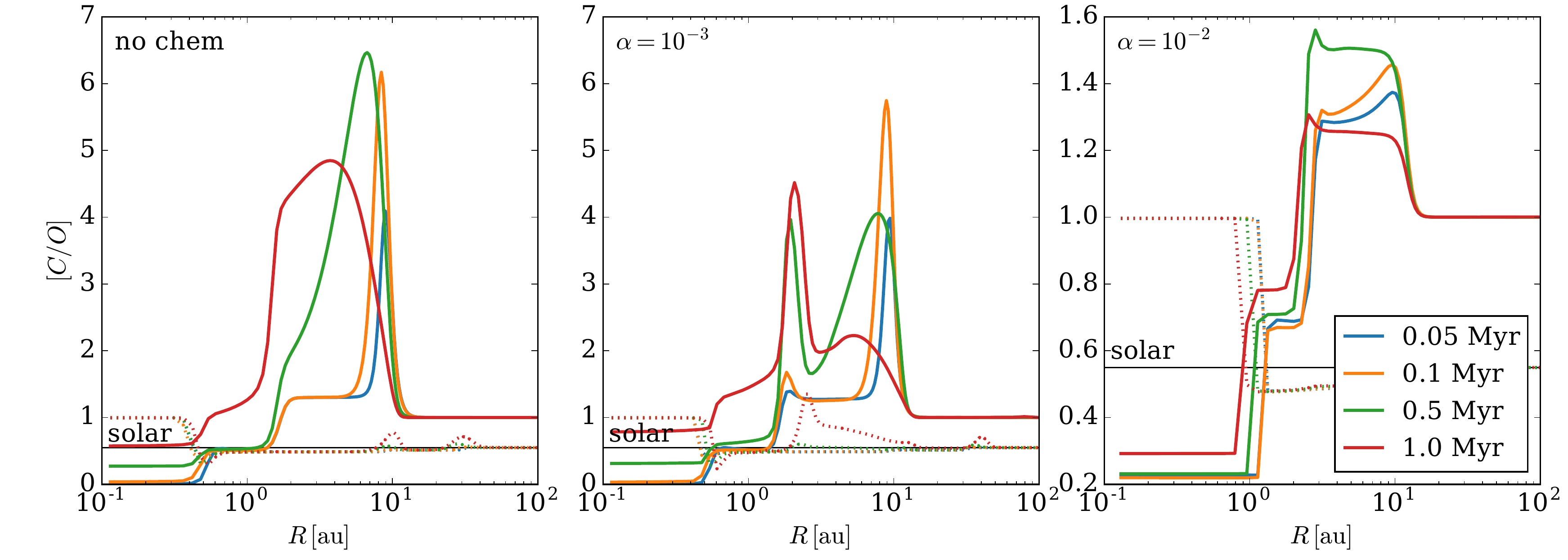}
    \caption{C/O ratio and carbon abundance evolution for models with radial drift. The carbon abundance shown is relative to the initial total carbon abundance ($1.38 \times 10^{-4}$, $0.3 \times {\rm solar}$). Note the different scales for the C/O ratio between the left and right panels.}
    \label{fig:drift_CHO}
\end{figure*}

The bulk abundances of the main atomic species, e.g. carbon, oxygen, and nitrogen, are of particular interest in the context of planet formation since the composition of exoplanet atmospheres must be connected to the competition of gas and solids in discs \citep[e.g.][]{oberg_2011, madhusudhan_2014}. We first explore the evolution of the C/O ratio in models without radial drift (\autoref{fig:visc_CO}). In these models, the total abundance (gas and ice phase) of carbon and oxygen bearing species does not change, only the partitioning of these species between the gas and ice phases.

\smallskip

The changes in composition driven by chemical processes are best seen in the model without transport (and $\alpha = 10^{-3}$). Inside the water ice line (0.5~au), all of the major carbon and oxygen are in the gas phase and thus chemical processing does not change the gas phase carbon or oxygen abundance. Between the water and CO$_2$ ice lines (1.5~au) the gas phase carbon and oxygen abundance is dominated by CO and CO$_2$, with a small contribution from hydrocarbons (CH$_4$ and C$_2$H$_2$ -- a proxy for C$_2$H$_6$; see \autoref{sec:results:chem:static}). Here, the C/O ratio increases as CO$_2$ is destroyed, producing CO and O, with O eventually forming  H$_2$O, which freezes out. Between the CO$_2$ and CH$_4$ ice lines (10~au), CO and CH$_4$ are initially the main gas phase carbon and oxygen carriers. CH$_4$ gets converted to larger hydrocarbons (C$_2$H$_2$), which freeze out causing the C/O ratio to reduce to 1. The spike in C/O ratio near the CO$_2$ ice line is due to the slightly lower binding energy of C$_2$H$_2$ than CO$_2$ (note that C$_2$H$_6$ also has a lower binding energy than CO$_2$, so \citealt{eistrup_2018} see a similar spike in C/O). Outside of  the CH$_4$ ice line CO is always the dominant the gas phase carbon and oxygen carrier, so the C/O ratio does not evolve. The nitrogen abundance evolution can be derived from \autoref{fig:N_chem} and is is slow, thus the C/N of N/O ratio evolution will be driven primarly by the evolution of carbon and oxygen, respectively.

\smallskip

As the speed of the viscous transport increases, the changes in composition driven by chemical reactions become restricted to regions of the disc at larger and larger radii. Already at $\alpha = 10^{-3}$ the effect of transport is first noticeable between the H$_2$O and CO$_2$ ice lines, since CO$_2$ no longer depletes \citep[see also][]{bosman_2018a}, while at $\alpha = 10^{-2}$ the carbon and oxygen abundance hardly evolve. At $\alpha = 10^{-2}$ the inward movement of the ice lines as the disc cools  generates the only significant changes in abundance ratio.

\smallskip

The influence of radial drift on the bulk abundances is sensitive to whether radial drift or viscous evolution is faster. When radial drift is included we see spikes in the gas phase C/H ratio at $\alpha = 10^{-3}$ (\autoref{fig:drift_CHO}), as reported by \citet{booth_2017}. These spikes are driven by volatile species being left at their snow lines when grains drift past them. We also see a peak in the C/O ratio inside of 10~au, which is driven by CH$_4$ entering the gas phase. Since radial drift depletes the disc of dust (and ice phase species) within a few 0.1~Myr the C/O ratio at 10~au begins to decline. This decline is due to a combination of the conversion of CH$_4$ to C$_2$H$_2$ and the viscous transport of CH$_4$ inwards from the ice line. After $\sim 0.5\,{\rm Myr}$, conversion of CH$_4$ to C$_2$H$_2$  generates a second spike in C/O ratio at a few au (where C$_2$H$_2$ enters the gas phase). At $\alpha = 10^{-2}$ the viscous and radial drift times are comparable, reducing the influence of radial drift. Nevertheless, we still find an enhancement in the gas phase C/H ratio by a factor 2 -- 3, since radial drift is still able to bring extra material into the inner regions. We again find C/O $> 1$ between the CH$_4$ and CO$_2$ snow lines since CH$_4$ is transported into this region faster than it is converted to C$_2$H$_2$. 

\section{Discussion}

\subsection{Impact of transport on disc composition}

In this paper we have investigated how the abundances of the main carbon, oxygen, and nitrogen carriers in the mid-plane of protoplanetary discs evolve under the competition between chemical kinetics and radial transport. The effects of transport differ between gas dominated transport (viscous evolution, and the transport of small grains at the viscous speed) and radial drift dominated transport. The main effect of gas dominated transport is to reduce the amount of time that a parcel of gas spends in regions where the processing occurs. Thus gas transport weakens the depletion of CO$_2$, CH$_4$ and CO from the gas phase that otherwise occurs inside their ice lines. Radial drift acts differently, enhancing the transport of molecular species inwards on grain surfaces. The molecules are then left at in the gas phase as they sublimate at their ice lines. Thus the main effect of radial drift is to enhance the abundances of molecular species inside their ice lines. Their subsequent evolution is controlled by transport further inward in the gas phase, along with processing by chemical reactions. A secondary effect of radial drift is that new species produced by chemical processing (e.g. C$_2$H$_2$ or H$_2$CO) can freeze out and be carried further in before re-entering the gas phase. 

\smallskip

The results presented here are relatively insensitive to the parameters in the model, except those that govern the rate of transport. In Appendix \ref{App:Params} we explore how the chemical depletion time-scales depend on model parameters in the absence of transport, showing that the results are only sensitive to the cosmic-ray ionization rate, $\zeta$, which facilitates the removal of CH$_4$ and CO from the gas phase. If $\zeta$ is lower than the canonical value of $10^{-17}$ (as suggested by \citealt{cleeves_2013}), then transport is expected to dominate for $\alpha \gtrsim 10^{-3}$, which is likely the case for most discs. Our results are otherwise insensitive to the disc mass or assumptions about the grain size distribution. This means that $\alpha$ and $\zeta$ essentially control the relative importance of transport and chemical reactions. There will also be a weak dependence on the temperature of the disc as the ice lines move to larger radius where the viscous time is longer.

\smallskip

The models presented here support the ideas put forward in \citet{booth_2017}, using similar assumptions for the transport of chemical species but simpler assumptions about their abundances. I.e. in models with efficient grain growth and radial drift the abundances inside ice lines are mostly controlled by the inward transport of molecular species from larger distances on grain surfaces, together with transport in the gas phase inside the ice line. Reducing the grain sizes weakens radial drift and results in the compositions being set largely by the composition at larger radii. The exception to this is when transport times are long ($\gg 1{\rm Myr}$ at $10\,{\rm au}$) and ionization rates are high ($\zeta \gtrsim 10^{-17}$), in which case chemical processing in the mid-plane may be able to remove CH$_4$ (and eventually CO) from the gas phase. We also verify the suggestion that, by processing molecules to less volatile species (such as CH$_4$ to C$_2$H$_2$), chemical reactions act to aid the transport of volatile species to smaller radii (\autoref{fig:drift_CHO}). 

\smallskip

The models presented here assume a turbulent viscosity, but it is possible that accretion in protoplanetary discs is primarily laminar, as discussed in \autoref{sec:methods:phys_params}. However, a low level of radial diffusion associated with weak turbulence would not affect the results presented here because the effects of radial diffusion are already localised to regions close to the snow lines. The primary effect of weak turbulence on the models is thus on grain growth and radial drift, which is limited by turbulent fragmentation in our $\alpha=0.01$ model. Lowering the strength of turbulence in that model by a factor 10 would result in efficient grain growth and radial drift, as in the $\alpha = 10^{-3}$ model. Reducing the strength of turbulence further would have little effect because grain growth becomes limited by radial drift instead.

A potentially more important effect associated with laminar accretion in discs is the possibility that the accretion occurs in a narrow layer far away from the mid-plane \citep[e.g.][]{gressel_2015}. In this case, vertical mixing is needed to couple the effects of accretion to the composition of the mid-plane. Two-dimensional models are needed to explore this in detail, but it is unlikely that the accretion flow can be completely decoupled from the mid-plane because vertical mixing should occur a factor of $(R/H)^2 \sim 10^2$--$10^3$ times faster than the radial mixing for the same $\alpha$. Thus an extraordinarily low $\alpha$ would be required to prevent vertical mixing altogether.

\smallskip

Dust traps, if widely present in protoplanetary discs, could have a significant impact on the abundance evolution. Such traps have typically been invoked to explain why dust grains can survive in discs for several Myr \citep[e.g.][]{pinilla_2012}. The impact of traps on the composition will depend on the evolution of the traps themselves. If the traps were to move with the gas, then the net effect would be similar to our models neglecting radial drift. However, if the traps remain stationary, preventing the inflow of dust entirely, the chemical evolution could be different as volatiles remain trapped at their original location. This would lead to the depletion of volatile species from the gas phase as the gas in the inner disc is accreted onto the star. This effect has been invoked to explain the depletion of refractory material accreting onto Herbig stars with transition discs \citep{kama_2015}.

\smallskip

Although the high CO abundance in our models appears in conflict with the observed depletion of CO in TW Hydra \citep{favre_2013, zhang_2017}, this can potentially be reconciled due to TW Hydra's age (8 Myr) and low accretion rate \citep{calvert_2002, donaldson_2016}, as long as the ionization rate is not too low ($\zeta \sim 10^{-17}$). In this case the viscous time-scale near the CO ice line is $\sim 5\,{\rm Myr}$, long enough that grain surface processes could deplete the CO abundance (requiring $\sim 3\,{\rm Myr}$, \citealt{bosman_2018b}). The low CO line fluxes observed for a range of discs \citep[e.g. those in Lupus;][]{miotello_2017} point to a more general depletion of CO within a few Myr. However, for systems requiring a higher $\alpha$ (e.g. HD 163296; \citealt{boneberg_2016}) or low ionization rate \citep{cleeves_2013}, grain surface processes are unlikely to be able to deplete CO rapidly enough.

\subsection{Implications for planet formation via pebble accretion}

The large difference in the gas-phase abundances in models with and without efficient radial drift (e.g. models with $\alpha = 10^{-3}$ and $10^{-2}$, respectively) suggests that exoplanet composition will be a powerful way to constrain planet formation by pebble accretion. Most models of giant planet formation by via pebble accretion assume large pebble sizes that drift efficiently \citep{lambrechts_2014, morbidelli_2015, bitsch_2015, bitsch_2019, ida_2016}. Planets forming in these models will have super-solar abundances unless they accrete their envelopes beyond the CO ice line \citep{booth_2017}. Our results suggest that these planets will likely also have ${\rm C/O} > 1$ and possibly ${\rm C/O} \gg 1$ if they accrete their envelopes inside the CH$_4$ ice line (which is the case in the models of \citealt{bitsch_2019}, for example). Recently, \citet{johansen_2018} suggested that giant planet formation via pebble accretion can occur for more modest pebble sizes, as long as the pebbles have settled to the mid-plane. Such models will be associated with more modest enhancements in the abundances, more similar to our $\alpha = 10^{-2}$ model (including radial drift).

\section{Conclusions}

We have explored the competition between chemical reactions in the mid-plane and transport due to viscous evolution and radial drift on the composition of protoplanetary discs. Our models show that the radial transport associated with accretion is able to transport material faster than it can be depleted due to chemical processing if the viscous $\alpha$ parameter is high ($\alpha \sim 10^{-2}$, \autoref{fig:static_chem}) or the ionization rate is low ($\zeta \sim 10^{-18}$, \autoref{fig:ion_evo}). For $\alpha = 10^{-3}$ and $\zeta = 10^{-17}$, transport will still overcome the depletion of CO$_2$ from the gas phase inside its ice line, but CO and CH$_4$ will remain depleted. 

\smallskip

Including radial drift enhances the gas phase abundances of the dominant molecular species inside their ice lines (\autoref{fig:drift_chem}). The strength of enhancement depends on how much faster radial drift is than transport in the gas. For commonly assumed parameters in models of dust evolution and planet formation pebble accretion \citep[e.g.]{birnstiel_2012, lambrechts_2014, bitsch_2015}, this leads to factor $\sim 10$ enhancements in the abundances. In models with more modest radial drift (grain sizes in the range 0.1--1~mm), radial drift can still lead to a factor 2--3 enhancement in the abundances inside 10~au.

\smallskip

In the coming years, the joint operation of the James Webb Space Telescope (JWST) and the Atacama Large Millimetre/submillimetre Array (ALMA) will enable robust measurements of the elemental composition of hot Jupiter atmospheres (such as C/O) alongside measurements of the chemical composition of planet forming discs.  Our results demonstrate the need to consider the interplay between both physical and chemical processes during protoplanetary disc evolution when interpretting such observations, with implications for the connection of planet composition to the composition of the disc from which they formed.

\section*{Acknowledgements}

We would like to thank Catherine Walsh, Cathie Clarke, and Mihkel Kama for useful discussions and careful reading of the manuscript. We additionally thank the referee, Alexander Cridland, for helpful suggestions which have improved this work.  RAB and JDI gratefully acknowledge support from the DISCSIM project, grant agreement 341137 under  ERC-2013-ADG.  JDI also acknowledges support from the STFC under ST/R000549/1. This project has made use of the SciPy stack \citep{Scipy}, including NumPy \citep{Numpy} and Matplotlib \citep{Matplotlib}.




\bibliographystyle{mnras}
\bibliography{references.bib} 

\begin{thebibliography}{}
\makeatletter
\relax
\def\mn@urlcharsother{\let\do\@makeother \do\$\do\&\do\#\do\^\do\_\do\%\do\~}
\def\mn@doi{\begingroup\mn@urlcharsother \@ifnextchar [ {\mn@doi@}
  {\mn@doi@[]}}
\def\mn@doi@[#1]#2{\def\@tempa{#1}\ifx\@tempa\@empty \href
  {http://dx.doi.org/#2} {doi:#2}\else \href {http://dx.doi.org/#2} {#1}\fi
  \endgroup}
\def\mn@eprint#1#2{\mn@eprint@#1:#2::\@nil}
\def\mn@eprint@arXiv#1{\href {http://arxiv.org/abs/#1} {{\tt arXiv:#1}}}
\def\mn@eprint@dblp#1{\href {http://dblp.uni-trier.de/rec/bibtex/#1.xml}
  {dblp:#1}}
\def\mn@eprint@#1:#2:#3:#4\@nil{\def\@tempa {#1}\def\@tempb {#2}\def\@tempc
  {#3}\ifx \@tempc \@empty \let \@tempc \@tempb \let \@tempb \@tempa \fi \ifx
  \@tempb \@empty \def\@tempb {arXiv}\fi \@ifundefined
  {mn@eprint@\@tempb}{\@tempb:\@tempc}{\expandafter \expandafter \csname
  mn@eprint@\@tempb\endcsname \expandafter{\@tempc}}}

\bibitem[\protect\citeauthoryear{{Aikawa}, {Umebayashi}, {Nakano}  \&
  {Miyama}}{{Aikawa} et~al.}{1999}]{aikawa_1999}
{Aikawa} Y.,  {Umebayashi} T.,  {Nakano} T.,   {Miyama} S.~M.,  1999, \mn@doi
  [\apj] {10.1086/307400}, \href
  {http://ukads.nottingham.ac.uk/abs/1999ApJ...519..705A} {519, 705}

\bibitem[\protect\citeauthoryear{{Ali-Dib}}{{Ali-Dib}}{2017}]{ali-dib_2017}
{Ali-Dib} M.,  2017, \mn@doi [\mnras] {10.1093/mnras/stw2651}, \href
  {http://adsabs.harvard.edu/abs/2017MNRAS.464.4282A} {464, 4282}

\bibitem[\protect\citeauthoryear{{Andrews} \& {Williams}}{{Andrews} \&
  {Williams}}{2007}]{andrews_2007}
{Andrews} S.~M.,  {Williams} J.~P.,  2007, \mn@doi [\apj] {10.1086/511741},
  \href {https://ui.adsabs.harvard.edu/\#abs/2007ApJ...659..705A} {659, 705}

\bibitem[\protect\citeauthoryear{{Ansdell} et~al.,}{{Ansdell}
  et~al.}{2016}]{ansdell_2016}
{Ansdell} M.,  et~al., 2016, \mn@doi [\apj] {10.3847/0004-637X/828/1/46}, \href
  {http://adsabs.harvard.edu/abs/2016ApJ...828...46A} {828, 46}

\bibitem[\protect\citeauthoryear{{Asplund}, {Grevesse}, {Sauval}  \&
  {Scott}}{{Asplund} et~al.}{2009}]{asplund_2009}
{Asplund} M.,  {Grevesse} N.,  {Sauval} A.~J.,   {Scott} P.,  2009, \mn@doi
  [Annual Review of Astronomy and Astrophysics]
  {10.1146/annurev.astro.46.060407.145222}, \href
  {https://ui.adsabs.harvard.edu/#abs/2009ARA&A..47..481A} {47, 481}

\bibitem[\protect\citeauthoryear{{Bai} \& {Stone}}{{Bai} \&
  {Stone}}{2013}]{bai_2013}
{Bai} X.-N.,  {Stone} J.~M.,  2013, \mn@doi [\apj]
  {10.1088/0004-637X/769/1/76}, \href
  {https://ui.adsabs.harvard.edu/\#abs/2013ApJ...769...76B} {769, 76}

\bibitem[\protect\citeauthoryear{{Bell} \& {Lin}}{{Bell} \&
  {Lin}}{1994}]{bell_1994}
{Bell} K.~R.,  {Lin} D.~N.~C.,  1994, \mn@doi [\apj] {10.1086/174206}, \href
  {https://ui.adsabs.harvard.edu/#abs/1994ApJ...427..987B} {427, 987}

\bibitem[\protect\citeauthoryear{{Bergin} et~al.,}{{Bergin}
  et~al.}{2013}]{bergin_2013}
{Bergin} E.~A.,  et~al., 2013, \mn@doi [\nat] {10.1038/nature11805}, \href
  {https://ui.adsabs.harvard.edu/\#abs/2013Natur.493..644B} {493, 644}

\bibitem[\protect\citeauthoryear{{Bergin}, {Du}, {Cleeves}, {Blake}, {Schwarz},
  {Visser}  \& {Zhang}}{{Bergin} et~al.}{2016}]{bergin_2016}
{Bergin} E.~A.,  {Du} F.,  {Cleeves} L.~I.,  {Blake} G.~A.,  {Schwarz} K.,
  {Visser} R.,   {Zhang} K.,  2016, \mn@doi [\apj]
  {10.3847/0004-637X/831/1/101}, \href
  {https://ui.adsabs.harvard.edu/\#abs/2016ApJ...831..101B} {831, 101}

\bibitem[\protect\citeauthoryear{{Birnstiel}, {Klahr}  \&
  {Ercolano}}{{Birnstiel} et~al.}{2012}]{birnstiel_2012}
{Birnstiel} T.,  {Klahr} H.,   {Ercolano} B.,  2012, \mn@doi [\aap]
  {10.1051/0004-6361/201118136}, \href
  {https://ui.adsabs.harvard.edu/#abs/2012A&A...539A.148B} {539, A148}

\bibitem[\protect\citeauthoryear{{Bitsch}, {Lambrechts}  \&
  {Johansen}}{{Bitsch} et~al.}{2015}]{bitsch_2015}
{Bitsch} B.,  {Lambrechts} M.,   {Johansen} A.,  2015, \mn@doi [\aap]
  {10.1051/0004-6361/201526463}, \href
  {https://ui.adsabs.harvard.edu/\#abs/2015A&A...582A.112B} {582, A112}

\bibitem[\protect\citeauthoryear{{Bitsch}, {Izidoro}, {Johansen}, {Raymond},
  {Morbidelli}, {Lambrechts}  \& {Jacobson}}{{Bitsch}
  et~al.}{2019}]{bitsch_2019}
{Bitsch} B.,  {Izidoro} A.,  {Johansen} A.,  {Raymond} S.~N.,  {Morbidelli} A.,
   {Lambrechts} M.,   {Jacobson} S.~A.,  2019, \mn@doi [\aap]
  {10.1051/0004-6361/201834489}, \href
  {https://ui.adsabs.harvard.edu/\#abs/2019A&A...623A..88B} {623, A88}

\bibitem[\protect\citeauthoryear{{Boneberg}, {Pani{\'c}}, {Haworth}, {Clarke}
  \& {Min}}{{Boneberg} et~al.}{2016}]{boneberg_2016}
{Boneberg} D.~M.,  {Pani{\'c}} O.,  {Haworth} T.~J.,  {Clarke} C.~J.,   {Min}
  M.,  2016, \mn@doi [\mnras] {10.1093/mnras/stw1325}, \href
  {http://adsabs.harvard.edu/abs/2016MNRAS.461..385B} {461, 385}

\bibitem[\protect\citeauthoryear{{Booth}, {Clarke}, {Madhusudhan}  \&
  {Ilee}}{{Booth} et~al.}{2017}]{booth_2017}
{Booth} R.~A.,  {Clarke} C.~J.,  {Madhusudhan} N.,   {Ilee} J.~D.,  2017,
  \mn@doi [\mnras] {10.1093/mnras/stx1103}, \href
  {https://ui.adsabs.harvard.edu/#abs/2017MNRAS.469.3994B} {469, 3994}

\bibitem[\protect\citeauthoryear{{Bosman}, {Tielens}  \& {van
  Dishoeck}}{{Bosman} et~al.}{2018a}]{bosman_2018a}
{Bosman} A.~D.,  {Tielens} A. G.~G.~M.,   {van Dishoeck} E.~F.,  2018a, \mn@doi
  [\aap] {10.1051/0004-6361/201732056}, \href
  {https://ui.adsabs.harvard.edu/#abs/2018A&A...611A..80B} {611, A80}

\bibitem[\protect\citeauthoryear{{Bosman}, {Walsh}  \& {van Dishoeck}}{{Bosman}
  et~al.}{2018b}]{bosman_2018b}
{Bosman} A.~D.,  {Walsh} C.,   {van Dishoeck} E.~F.,  2018b, \mn@doi [\aap]
  {10.1051/0004-6361/201833497}, \href
  {https://ui.adsabs.harvard.edu/#abs/2018A&A...618A.182B} {618, A182}

\bibitem[\protect\citeauthoryear{{Calvet}, {D'Alessio}, {Hartmann}, {Wilner},
  {Walsh}  \& {Sitko}}{{Calvet} et~al.}{2002}]{calvert_2002}
{Calvet} N.,  {D'Alessio} P.,  {Hartmann} L.,  {Wilner} D.,  {Walsh} A.,
  {Sitko} M.,  2002, \mn@doi [\apj] {10.1086/339061}, \href
  {http://adsabs.harvard.edu/abs/2002ApJ...568.1008C} {568, 1008}

\bibitem[\protect\citeauthoryear{{Clarke} et~al.,}{{Clarke}
  et~al.}{2018}]{clarke_2018}
{Clarke} C.~J.,  et~al., 2018, \mn@doi [\apjl] {10.3847/2041-8213/aae36b},
  \href {http://adsabs.harvard.edu/abs/2018ApJ...866L...6C} {866, L6}

\bibitem[\protect\citeauthoryear{{Cleeves}, {Adams}  \& {Bergin}}{{Cleeves}
  et~al.}{2013a}]{cleeves_2013}
{Cleeves} L.~I.,  {Adams} F.~C.,   {Bergin} E.~A.,  2013a, \mn@doi [\apj]
  {10.1088/0004-637X/772/1/5}, \href
  {https://ui.adsabs.harvard.edu/\#abs/2013ApJ...772....5C} {772, 5}

\bibitem[\protect\citeauthoryear{{Cleeves}, {Adams}, {Bergin}  \&
  {Visser}}{{Cleeves} et~al.}{2013b}]{cleeves_2013b}
{Cleeves} L.~I.,  {Adams} F.~C.,  {Bergin} E.~A.,   {Visser} R.,  2013b,
  \mn@doi [\apj] {10.1088/0004-637X/777/1/28}, \href
  {https://ui.adsabs.harvard.edu/\#abs/2013ApJ...777...28C} {777, 28}

\bibitem[\protect\citeauthoryear{{Collings}, {Anderson}, {Chen}, {Dever},
  {Viti}, {Williams}  \& {McCoustra}}{{Collings} et~al.}{2004}]{collings_2004}
{Collings} M.~P.,  {Anderson} M.~A.,  {Chen} R.,  {Dever} J.~W.,  {Viti} S.,
  {Williams} D.~A.,   {McCoustra} M.~R.~S.,  2004, \mn@doi [\mnras]
  {10.1111/j.1365-2966.2004.08272.x}, \href
  {http://ukads.nottingham.ac.uk/abs/2004MNRAS.354.1133C} {354, 1133}

\bibitem[\protect\citeauthoryear{{Cridland}, {Pudritz}  \& {Alessi}}{{Cridland}
  et~al.}{2016}]{cridland_2016}
{Cridland} A.~J.,  {Pudritz} R.~E.,   {Alessi} M.,  2016, \mn@doi [\mnras]
  {10.1093/mnras/stw1511}, \href
  {https://ui.adsabs.harvard.edu/#abs/2016MNRAS.461.3274C} {461, 3274}

\bibitem[\protect\citeauthoryear{{Cridland}, {Pudritz}, {Birnstiel}, {Cleeves}
  \& {Bergin}}{{Cridland} et~al.}{2017}]{cridland_2017}
{Cridland} A.~J.,  {Pudritz} R.~E.,  {Birnstiel} T.,  {Cleeves} L.~I.,
  {Bergin} E.~A.,  2017, \mn@doi [\mnras] {10.1093/mnras/stx1069}, \href
  {https://ui.adsabs.harvard.edu/#abs/2017MNRAS.469.3910C} {469, 3910}

\bibitem[\protect\citeauthoryear{{Donaldson}, {Weinberger}, {Gagn{\'e}},
  {Faherty}, {Boss}  \& {Keiser}}{{Donaldson} et~al.}{2016}]{donaldson_2016}
{Donaldson} J.~K.,  {Weinberger} A.~J.,  {Gagn{\'e}} J.,  {Faherty} J.~K.,
  {Boss} A.~P.,   {Keiser} S.~A.,  2016, \mn@doi [\apj]
  {10.3847/1538-4357/833/1/95}, \href
  {http://adsabs.harvard.edu/abs/2016ApJ...833...95D} {833, 95}

\bibitem[\protect\citeauthoryear{{Eistrup}, {Walsh}  \& {van
  Dishoeck}}{{Eistrup} et~al.}{2016}]{eistrup_2016}
{Eistrup} C.,  {Walsh} C.,   {van Dishoeck} E.~F.,  2016, \mn@doi [\aap]
  {10.1051/0004-6361/201628509}, \href
  {http://ukads.nottingham.ac.uk/abs/2016A%26A...595A..83E} {595, A83}

\bibitem[\protect\citeauthoryear{{Eistrup}, {Walsh}  \& {van
  Dishoeck}}{{Eistrup} et~al.}{2018}]{eistrup_2018}
{Eistrup} C.,  {Walsh} C.,   {van Dishoeck} E.~F.,  2018, \mn@doi [\aap]
  {10.1051/0004-6361/201731302}, \href
  {https://ui.adsabs.harvard.edu/#abs/2018A&A...613A..14E} {613, A14}

\bibitem[\protect\citeauthoryear{{Ercolano}, {Rosotti}, {Picogna}  \&
  {Testi}}{{Ercolano} et~al.}{2017}]{ercolano_2017}
{Ercolano} B.,  {Rosotti} G.~P.,  {Picogna} G.,   {Testi} L.,  2017, \mn@doi
  [\mnras] {10.1093/mnrasl/slw188}, \href
  {https://ui.adsabs.harvard.edu/\#abs/2017MNRAS.464L..95E} {464, L95}

\bibitem[\protect\citeauthoryear{{Evans}, {Ilee}, {Boley}, {Caselli},
  {Durisen}, {Hartquist}  \& {Rawlings}}{{Evans} et~al.}{2015}]{evans_2015}
{Evans} M.~G.,  {Ilee} J.~D.,  {Boley} A.~C.,  {Caselli} P.,  {Durisen} R.~H.,
  {Hartquist} T.~W.,   {Rawlings} J.~M.~C.,  2015, \mn@doi [\mnras]
  {10.1093/mnras/stv1698}, \href
  {http://adsabs.harvard.edu/abs/2015MNRAS.453.1147E} {453, 1147}

\bibitem[\protect\citeauthoryear{{Favre}, {Cleeves}, {Bergin}, {Qi}  \&
  {Blake}}{{Favre} et~al.}{2013}]{favre_2013}
{Favre} C.,  {Cleeves} L.~I.,  {Bergin} E.~A.,  {Qi} C.,   {Blake} G.~A.,
  2013, \mn@doi [\apj] {10.1088/2041-8205/776/2/L38}, \href
  {https://ui.adsabs.harvard.edu/\#abs/2013ApJ...776L..38F} {776, L38}

\bibitem[\protect\citeauthoryear{{Flaherty} et~al.,}{{Flaherty}
  et~al.}{2017}]{flaherty_2017}
{Flaherty} K.~M.,  et~al., 2017, \mn@doi [\apj] {10.3847/1538-4357/aa79f9},
  \href {http://adsabs.harvard.edu/abs/2017ApJ...843..150F} {843, 150}

\bibitem[\protect\citeauthoryear{{Flaherty}, {Hughes}, {Teague}, {Simon},
  {Andrews}  \& {Wilner}}{{Flaherty} et~al.}{2018}]{flaherty_2018}
{Flaherty} K.~M.,  {Hughes} A.~M.,  {Teague} R.,  {Simon} J.~B.,  {Andrews}
  S.~M.,   {Wilner} D.~J.,  2018, \mn@doi [\apj] {10.3847/1538-4357/aab615},
  \href {http://adsabs.harvard.edu/abs/2018ApJ...856..117F} {856, 117}

\bibitem[\protect\citeauthoryear{{Fromang}, {Latter}, {Lesur}  \&
  {Ogilvie}}{{Fromang} et~al.}{2013}]{fromang_2013}
{Fromang} S.,  {Latter} H.,  {Lesur} G.,   {Ogilvie} G.~I.,  2013, \mn@doi
  [\aap] {10.1051/0004-6361/201220016}, \href
  {https://ui.adsabs.harvard.edu/\#abs/2013A&A...552A..71F} {552, A71}

\bibitem[\protect\citeauthoryear{{Gail} \& {Trieloff}}{{Gail} \&
  {Trieloff}}{2017}]{gail_2017}
{Gail} H.-P.,  {Trieloff} M.,  2017, \mn@doi [\aap]
  {10.1051/0004-6361/201730480}, \href
  {https://ui.adsabs.harvard.edu/\#abs/2017A&A...606A..16G} {606, A16}

\bibitem[\protect\citeauthoryear{{Grassi}, {Bovino}, {Schleicher}, {Prieto},
  {Seifried}, {Simoncini}  \& {Gianturco}}{{Grassi} et~al.}{2014}]{grassi_2014}
{Grassi} T.,  {Bovino} S.,  {Schleicher} D.~R.~G.,  {Prieto} J.,  {Seifried}
  D.,  {Simoncini} E.,   {Gianturco} F.~A.,  2014, \mn@doi [\mnras]
  {10.1093/mnras/stu114}, \href
  {http://adsabs.harvard.edu/abs/2014MNRAS.439.2386G} {439, 2386}

\bibitem[\protect\citeauthoryear{{Greaves} \& {Rice}}{{Greaves} \&
  {Rice}}{2010}]{greaves_2010}
{Greaves} J.~S.,  {Rice} W.~K.~M.,  2010, \mn@doi [\mnras]
  {10.1111/j.1365-2966.2010.17043.x}, \href
  {https://ui.adsabs.harvard.edu/\#abs/2010MNRAS.407.1981G} {407, 1981}

\bibitem[\protect\citeauthoryear{{Gressel}, {Turner}, {Nelson}  \&
  {McNally}}{{Gressel} et~al.}{2015}]{gressel_2015}
{Gressel} O.,  {Turner} N.~J.,  {Nelson} R.~P.,   {McNally} C.~P.,  2015,
  \mn@doi [\apj] {10.1088/0004-637X/801/2/84}, \href
  {https://ui.adsabs.harvard.edu/abs/2015ApJ...801...84G} {801, 84}

\bibitem[\protect\citeauthoryear{{Guilloteau}, {Dutrey}, {Pi{\'e}tu}  \&
  {Boehler}}{{Guilloteau} et~al.}{2011}]{guilloteau_2011}
{Guilloteau} S.,  {Dutrey} A.,  {Pi{\'e}tu} V.,   {Boehler} Y.,  2011, \mn@doi
  [\aap] {10.1051/0004-6361/201015209}, \href
  {https://ui.adsabs.harvard.edu/\#abs/2011A&A...529A.105G} {529, A105}

\bibitem[\protect\citeauthoryear{{Heays}, {Bosman}  \& {van Dishoeck}}{{Heays}
  et~al.}{2017}]{heays_2017}
{Heays} A.~N.,  {Bosman} A.~D.,   {van Dishoeck} E.~F.,  2017, \mn@doi [\aap]
  {10.1051/0004-6361/201628742}, \href
  {https://ui.adsabs.harvard.edu/\#abs/2017A&A...602A.105H} {602, A105}

\bibitem[\protect\citeauthoryear{{Heinzeller}, {Nomura}, {Walsh}  \&
  {Millar}}{{Heinzeller} et~al.}{2011}]{heinzeller_2011}
{Heinzeller} D.,  {Nomura} H.,  {Walsh} C.,   {Millar} T.~J.,  2011, \mn@doi
  [\apj] {10.1088/0004-637X/731/2/115}, \href
  {https://ui.adsabs.harvard.edu/\#abs/2011ApJ...731..115H} {731, 115}

\bibitem[\protect\citeauthoryear{{Henning} \& {Semenov}}{{Henning} \&
  {Semenov}}{2013}]{henning_2013}
{Henning} T.,  {Semenov} D.,  2013, \mn@doi [Chemical Reviews]
  {10.1021/cr400128p}, \href
  {http://adsabs.harvard.edu/abs/2013ChRv..113.9016H} {113, 9016}

\bibitem[\protect\citeauthoryear{{Hindmarsh}}{{Hindmarsh}}{1983}]{hindmarsh_1983}
{Hindmarsh} A.~C.,  1983, IMACS Transactions on Scientific Computation, 1, 55

\bibitem[\protect\citeauthoryear{Hunter}{Hunter}{2007}]{Matplotlib}
Hunter J.~D.,  2007, \mn@doi [Computing in Science Engineering]
  {10.1109/MCSE.2007.55}, 9, 90

\bibitem[\protect\citeauthoryear{{Ida}, {Guillot}  \& {Morbidelli}}{{Ida}
  et~al.}{2016}]{ida_2016}
{Ida} S.,  {Guillot} T.,   {Morbidelli} A.,  2016, \mn@doi [\aap]
  {10.1051/0004-6361/201628099}, \href
  {https://ui.adsabs.harvard.edu/\#abs/2016A&A...591A..72I} {591, A72}

\bibitem[\protect\citeauthoryear{{Ilee}, {Boley}, {Caselli}, {Durisen},
  {Hartquist}  \& {Rawlings}}{{Ilee} et~al.}{2011}]{ilee_2011}
{Ilee} J.~D.,  {Boley} A.~C.,  {Caselli} P.,  {Durisen} R.~H.,  {Hartquist}
  T.~W.,   {Rawlings} J.~M.~C.,  2011, \mn@doi [\mnras]
  {10.1111/j.1365-2966.2011.19455.x}, \href
  {http://adsabs.harvard.edu/abs/2011MNRAS.417.2950I} {417, 2950}

\bibitem[\protect\citeauthoryear{{Ilee} et~al.,}{{Ilee}
  et~al.}{2017}]{ilee_2017}
{Ilee} J.~D.,  et~al., 2017, \mn@doi [\mnras] {10.1093/mnras/stx1966}, \href
  {http://adsabs.harvard.edu/abs/2017MNRAS.472..189I} {472, 189}

\bibitem[\protect\citeauthoryear{{Johansen}, {Ida}  \& {Brasser}}{{Johansen}
  et~al.}{2018}]{johansen_2018}
{Johansen} A.,  {Ida} S.,   {Brasser} R.,  2018, arXiv e-prints, \href
  {https://ui.adsabs.harvard.edu/\#abs/2018arXiv181100523J} {p.
  arXiv:1811.00523}

\bibitem[\protect\citeauthoryear{Jones, Oliphant, Peterson  et~al.}{Jones
  et~al.}{2001}]{Scipy}
Jones E.,  Oliphant T.,  Peterson P.,   et~al., 2001, {SciPy}: Open source
  scientific tools for {Python}, \url {http://www.scipy.org/}

\bibitem[\protect\citeauthoryear{{Kama}, {Folsom}  \& {Pinilla}}{{Kama}
  et~al.}{2015}]{kama_2015}
{Kama} M.,  {Folsom} C.~P.,   {Pinilla} P.,  2015, \mn@doi [\aap]
  {10.1051/0004-6361/201527094}, \href
  {https://ui.adsabs.harvard.edu/abs/2015A&A...582L..10K} {582, L10}

\bibitem[\protect\citeauthoryear{{Kama} et~al.,}{{Kama}
  et~al.}{2016}]{kama_2016}
{Kama} M.,  et~al., 2016, \mn@doi [\aap] {10.1051/0004-6361/201526991}, \href
  {https://ui.adsabs.harvard.edu/\#abs/2016A&A...592A..83K} {592, A83}

\bibitem[\protect\citeauthoryear{{Klarmann}, {Ormel}  \& {Dominik}}{{Klarmann}
  et~al.}{2018}]{klarmann_2018}
{Klarmann} L.,  {Ormel} C.~W.,   {Dominik} C.,  2018, \mn@doi [\aap]
  {10.1051/0004-6361/201833719}, \href
  {https://ui.adsabs.harvard.edu/\#abs/2018A&A...618L...1K} {618, L1}

\bibitem[\protect\citeauthoryear{{Krijt}, {Schwarz}, {Bergin}  \&
  {Ciesla}}{{Krijt} et~al.}{2018}]{krijt_2018}
{Krijt} S.,  {Schwarz} K.~R.,  {Bergin} E.~A.,   {Ciesla} F.~J.,  2018, \mn@doi
  [\apj] {10.3847/1538-4357/aad69b}, \href
  {https://ui.adsabs.harvard.edu/\#abs/2018ApJ...864...78K} {864, 78}

\bibitem[\protect\citeauthoryear{{Lambrechts} \& {Johansen}}{{Lambrechts} \&
  {Johansen}}{2012}]{lambrechts_2012}
{Lambrechts} M.,  {Johansen} A.,  2012, \mn@doi [\aap]
  {10.1051/0004-6361/201219127}, \href
  {https://ui.adsabs.harvard.edu/\#abs/2012A&A...544A..32L} {544, A32}

\bibitem[\protect\citeauthoryear{{Lambrechts} \& {Johansen}}{{Lambrechts} \&
  {Johansen}}{2014}]{lambrechts_2014}
{Lambrechts} M.,  {Johansen} A.,  2014, \mn@doi [\aap]
  {10.1051/0004-6361/201424343}, \href
  {https://ui.adsabs.harvard.edu/\#abs/2014A&A...572A.107L} {572, A107}

\bibitem[\protect\citeauthoryear{{Lodato}, {Scardoni}, {Manara}  \&
  {Testi}}{{Lodato} et~al.}{2017}]{lodato_2017}
{Lodato} G.,  {Scardoni} C.~E.,  {Manara} C.~F.,   {Testi} L.,  2017, \mn@doi
  [\mnras] {10.1093/mnras/stx2273}, \href
  {https://ui.adsabs.harvard.edu/\#abs/2017MNRAS.472.4700L} {472, 4700}

\bibitem[\protect\citeauthoryear{{Lodders}}{{Lodders}}{2010}]{lodders_2010}
{Lodders} K.,  2010, {Exoplanet Chemistry}.
Wiley, p.~157, \mn@doi{10.1002/9783527629763.ch8}

\bibitem[\protect\citeauthoryear{{Lynden-Bell} \& {Pringle}}{{Lynden-Bell} \&
  {Pringle}}{1974}]{1ynden-bell_1974}
{Lynden-Bell} D.,  {Pringle} J.~E.,  1974, \mn@doi [\mnras]
  {10.1093/mnras/168.3.603}, \href
  {https://ui.adsabs.harvard.edu/#abs/1974MNRAS.168..603L} {168, 603}

\bibitem[\protect\citeauthoryear{{Madhusudhan}}{{Madhusudhan}}{2012}]{madhusudhan_2012}
{Madhusudhan} N.,  2012, \mn@doi [\apj] {10.1088/0004-637X/758/1/36}, \href
  {https://ui.adsabs.harvard.edu/\#abs/2012ApJ...758...36M} {758, 36}

\bibitem[\protect\citeauthoryear{{Madhusudhan}, {Amin}  \&
  {Kennedy}}{{Madhusudhan} et~al.}{2014}]{madhusudhan_2014}
{Madhusudhan} N.,  {Amin} M.~A.,   {Kennedy} G.~M.,  2014, \mn@doi [\apjl]
  {10.1088/2041-8205/794/1/L12}, \href
  {http://adsabs.harvard.edu/abs/2014ApJ...794L..12M} {794, L12}

\bibitem[\protect\citeauthoryear{{Madhusudhan}, {Bitsch}, {Johansen}  \&
  {Eriksson}}{{Madhusudhan} et~al.}{2017}]{madhusudhan_2017}
{Madhusudhan} N.,  {Bitsch} B.,  {Johansen} A.,   {Eriksson} L.,  2017, \mn@doi
  [\mnras] {10.1093/mnras/stx1139}, \href
  {https://ui.adsabs.harvard.edu/\#abs/2017MNRAS.469.4102M} {469, 4102}

\bibitem[\protect\citeauthoryear{{Manara} et~al.,}{{Manara}
  et~al.}{2016}]{manara_2016}
{Manara} C.~F.,  et~al., 2016, \mn@doi [\aap] {10.1051/0004-6361/201628549},
  \href {https://ui.adsabs.harvard.edu/\#abs/2016A&A...591L...3M} {591, L3}

\bibitem[\protect\citeauthoryear{{Manara}, {Morbidelli}  \& {Guillot}}{{Manara}
  et~al.}{2018}]{manara_2018}
{Manara} C.~F.,  {Morbidelli} A.,   {Guillot} T.,  2018, \mn@doi [\aap]
  {10.1051/0004-6361/201834076}, \href
  {https://ui.adsabs.harvard.edu/\#abs/2018A&A...618L...3M} {618, L3}

\bibitem[\protect\citeauthoryear{{Millar}, {Farquhar}  \& {Willacy}}{{Millar}
  et~al.}{1997}]{millar_umist_1997}
{Millar} T.~J.,  {Farquhar} P.~R.~A.,   {Willacy} K.,  1997, \mn@doi [\aaps]
  {10.1051/aas:1997118}, \href
  {http://ukads.nottingham.ac.uk/abs/1997A%26AS..121..139M} {121, 139}

\bibitem[\protect\citeauthoryear{{Miotello} et~al.,}{{Miotello}
  et~al.}{2017}]{miotello_2017}
{Miotello} A.,  et~al., 2017, \mn@doi [\aap] {10.1051/0004-6361/201629556},
  \href {https://ui.adsabs.harvard.edu/\#abs/2017A&A...599A.113M} {599, A113}

\bibitem[\protect\citeauthoryear{{Morbidelli}, {Lambrechts}, {Jacobson}  \&
  {Bitsch}}{{Morbidelli} et~al.}{2015}]{morbidelli_2015}
{Morbidelli} A.,  {Lambrechts} M.,  {Jacobson} S.,   {Bitsch} B.,  2015,
  \mn@doi [\icarus] {10.1016/j.icarus.2015.06.003}, \href
  {https://ui.adsabs.harvard.edu/\#abs/2015Icar..258..418M} {258, 418}

\bibitem[\protect\citeauthoryear{{Mordasini}, {van Boekel}, {Molli{\`e}re},
  {Henning}  \& {Benneke}}{{Mordasini} et~al.}{2016}]{mordasini_2016}
{Mordasini} C.,  {van Boekel} R.,  {Molli{\`e}re} P.,  {Henning} T.,
  {Benneke} B.,  2016, \mn@doi [\apj] {10.3847/0004-637X/832/1/41}, \href
  {https://ui.adsabs.harvard.edu/\#abs/2016ApJ...832...41M} {832, 41}

\bibitem[\protect\citeauthoryear{{Mulders}, {Pascucci}, {Manara}, {Testi},
  {Herczeg}, {Henning}, {Mohanty}  \& {Lodato}}{{Mulders}
  et~al.}{2017}]{mulders_2017}
{Mulders} G.~D.,  {Pascucci} I.,  {Manara} C.~F.,  {Testi} L.,  {Herczeg}
  G.~J.,  {Henning} T.,  {Mohanty} S.,   {Lodato} G.,  2017, \mn@doi [\apj]
  {10.3847/1538-4357/aa8906}, \href
  {https://ui.adsabs.harvard.edu/\#abs/2017ApJ...847...31M} {847, 31}

\bibitem[\protect\citeauthoryear{{Mumma} \& {Charnley}}{{Mumma} \&
  {Charnley}}{2011}]{mumma_2011}
{Mumma} M.~J.,  {Charnley} S.~B.,  2011, \mn@doi [Annual Review of Astronomy
  and Astrophysics] {10.1146/annurev-astro-081309-130811}, \href
  {https://ui.adsabs.harvard.edu/\#abs/2011ARA&A..49..471M} {49, 471}

\bibitem[\protect\citeauthoryear{{Najita} \& {Kenyon}}{{Najita} \&
  {Kenyon}}{2014}]{najita_2014}
{Najita} J.~R.,  {Kenyon} S.~J.,  2014, \mn@doi [\mnras]
  {10.1093/mnras/stu1994}, \href
  {https://ui.adsabs.harvard.edu/\#abs/2014MNRAS.445.3315N} {445, 3315}

\bibitem[\protect\citeauthoryear{{Nomura}, {Aikawa}, {Nakagawa}  \&
  {Millar}}{{Nomura} et~al.}{2009}]{nomura_2009}
{Nomura} H.,  {Aikawa} Y.,  {Nakagawa} Y.,   {Millar} T.~J.,  2009, \mn@doi
  [\aap] {10.1051/0004-6361:200810206}, \href
  {https://ui.adsabs.harvard.edu/\#abs/2009A&A...495..183N} {495, 183}

\bibitem[\protect\citeauthoryear{{{\"O}berg}, {Garrod}, {van Dishoeck}  \&
  {Linnartz}}{{{\"O}berg} et~al.}{2009}]{oberg_2009}
{{\"O}berg} K.~I.,  {Garrod} R.~T.,  {van Dishoeck} E.~F.,   {Linnartz} H.,
  2009, \mn@doi [\aap] {10.1051/0004-6361/200912559}, \href
  {http://ukads.nottingham.ac.uk/abs/2009A%26A...504..891O} {504, 891}

\bibitem[\protect\citeauthoryear{{{\"O}berg}, {Murray-Clay}  \&
  {Bergin}}{{{\"O}berg} et~al.}{2011}]{oberg_2011}
{{\"O}berg} K.~I.,  {Murray-Clay} R.,   {Bergin} E.~A.,  2011, \mn@doi [\apjl]
  {10.1088/2041-8205/743/1/L16}, \href
  {http://adsabs.harvard.edu/abs/2011ApJ...743L..16O} {743, L16}

\bibitem[\protect\citeauthoryear{Oliphant}{Oliphant}{2007}]{Numpy}
Oliphant T.~E.,  2007, \mn@doi [Computing in Science Engineering]
  {10.1109/MCSE.2007.58}, 9, 10

\bibitem[\protect\citeauthoryear{{Ormel} \& {Klahr}}{{Ormel} \&
  {Klahr}}{2010}]{ormel_2010}
{Ormel} C.~W.,  {Klahr} H.~H.,  2010, \mn@doi [\aap]
  {10.1051/0004-6361/201014903}, \href
  {https://ui.adsabs.harvard.edu/\#abs/2010A&A...520A..43O} {520, A43}

\bibitem[\protect\citeauthoryear{{Penteado}, {Walsh}  \& {Cuppen}}{{Penteado}
  et~al.}{2017}]{penteado_2017}
{Penteado} E.~M.,  {Walsh} C.,   {Cuppen} H.~M.,  2017, \mn@doi [\apj]
  {10.3847/1538-4357/aa78f9}, \href
  {http://ukads.nottingham.ac.uk/abs/2017ApJ...844...71P} {844, 71}

\bibitem[\protect\citeauthoryear{{Pinilla}, {Birnstiel}, {Ricci}, {Dullemond},
  {Uribe}, {Testi}  \& {Natta}}{{Pinilla} et~al.}{2012}]{pinilla_2012}
{Pinilla} P.,  {Birnstiel} T.,  {Ricci} L.,  {Dullemond} C.~P.,  {Uribe} A.~L.,
   {Testi} L.,   {Natta} A.,  2012, \mn@doi [\aap]
  {10.1051/0004-6361/201118204}, \href
  {https://ui.adsabs.harvard.edu/\#abs/2012A&A...538A.114P} {538, A114}

\bibitem[\protect\citeauthoryear{{Pinte}, {Dent}, {M{\'e}nard}, {Hales},
  {Hill}, {Cortes}  \& {de Gregorio-Monsalvo}}{{Pinte}
  et~al.}{2016}]{pinte_2016}
{Pinte} C.,  {Dent} W.~R.~F.,  {M{\'e}nard} F.,  {Hales} A.,  {Hill} T.,
  {Cortes} P.,   {de Gregorio-Monsalvo} I.,  2016, \mn@doi [\apj]
  {10.3847/0004-637X/816/1/25}, \href
  {https://ui.adsabs.harvard.edu/\#abs/2016ApJ...816...25P} {816, 25}

\bibitem[\protect\citeauthoryear{{Piso}, {{\"O}berg}, {Birnstiel}  \&
  {Murray-Clay}}{{Piso} et~al.}{2015}]{piso_2015}
{Piso} A.-M.~A.,  {{\"O}berg} K.~I.,  {Birnstiel} T.,   {Murray-Clay} R.~A.,
  2015, \mn@doi [\apj] {10.1088/0004-637X/815/2/109}, \href
  {https://ui.adsabs.harvard.edu/\#abs/2015ApJ...815..109P} {815, 109}

\bibitem[\protect\citeauthoryear{{Piso}, {Pegues}  \& {{\"O}berg}}{{Piso}
  et~al.}{2016}]{piso_2016}
{Piso} A.-M.~A.,  {Pegues} J.,   {{\"O}berg} K.~I.,  2016, \mn@doi [\apj]
  {10.3847/1538-4357/833/2/203}, \href
  {https://ui.adsabs.harvard.edu/\#abs/2016ApJ...833..203P} {833, 203}

\bibitem[\protect\citeauthoryear{{Pollack}, {Hollenbach}, {Beckwith},
  {Simonelli}, {Roush}  \& {Fong}}{{Pollack} et~al.}{1994}]{pollack_1994}
{Pollack} J.~B.,  {Hollenbach} D.,  {Beckwith} S.,  {Simonelli} D.~P.,  {Roush}
  T.,   {Fong} W.,  1994, \mn@doi [\apj] {10.1086/173677}, \href
  {https://ui.adsabs.harvard.edu/#abs/1994ApJ...421..615P} {421, 615}

\bibitem[\protect\citeauthoryear{{Pontoppidan}, {Salyk}, {Banzatti}, {Blake},
  {Walsh}, {Lacy}  \& {Richter}}{{Pontoppidan} et~al.}{2019}]{pontoppidan_2019}
{Pontoppidan} K.~M.,  {Salyk} C.,  {Banzatti} A.,  {Blake} G.~A.,  {Walsh} C.,
  {Lacy} J.~H.,   {Richter} M.~J.,  2019, arXiv e-prints, \href
  {https://ui.adsabs.harvard.edu/\#abs/2019arXiv190203647P} {p.
  arXiv:1902.03647}

\bibitem[\protect\citeauthoryear{{Rafikov}}{{Rafikov}}{2017}]{rafikov_2017}
{Rafikov} R.~R.,  2017, \mn@doi [\apj] {10.3847/1538-4357/aa6249}, \href
  {https://ui.adsabs.harvard.edu/\#abs/2017ApJ...837..163R} {837, 163}

\bibitem[\protect\citeauthoryear{{Schwarz} \& {Bergin}}{{Schwarz} \&
  {Bergin}}{2014}]{schwarz_2014}
{Schwarz} K.~R.,  {Bergin} E.~A.,  2014, \mn@doi [\apj]
  {10.1088/0004-637X/797/2/113}, \href
  {https://ui.adsabs.harvard.edu/\#abs/2014ApJ...797..113S} {797, 113}

\bibitem[\protect\citeauthoryear{{Semenov} \& {Wiebe}}{{Semenov} \&
  {Wiebe}}{2011}]{semenov_2011}
{Semenov} D.,  {Wiebe} D.,  2011, \mn@doi [The Astrophysical Journal Supplement
  Series] {10.1088/0067-0049/196/2/25}, \href
  {https://ui.adsabs.harvard.edu/\#abs/2011ApJS..196...25S} {196, 25}

\bibitem[\protect\citeauthoryear{{Stammler}, {Birnstiel}, {Pani{\'c}},
  {Dullemond}  \& {Dominik}}{{Stammler} et~al.}{2017}]{stammler_2017}
{Stammler} S.~M.,  {Birnstiel} T.,  {Pani{\'c}} O.,  {Dullemond} C.~P.,
  {Dominik} C.,  2017, \mn@doi [\aap] {10.1051/0004-6361/201629041}, \href
  {https://ui.adsabs.harvard.edu/#abs/2017A&A...600A.140S} {600, A140}

\bibitem[\protect\citeauthoryear{{Suzuki} \& {Inutsuka}}{{Suzuki} \&
  {Inutsuka}}{2009}]{suzuki_2009}
{Suzuki} T.~K.,  {Inutsuka} S.-i.,  2009, \mn@doi [\apj]
  {10.1088/0004-637X/691/1/L49}, \href
  {https://ui.adsabs.harvard.edu/\#abs/2009ApJ...691L..49S} {691, L49}

\bibitem[\protect\citeauthoryear{{Takeuchi} \& {Lin}}{{Takeuchi} \&
  {Lin}}{2002}]{takeuchi_2002}
{Takeuchi} T.,  {Lin} D.~N.~C.,  2002, \mn@doi [\apj] {10.1086/344437}, \href
  {https://ui.adsabs.harvard.edu/\#abs/2002ApJ...581.1344T} {581, 1344}

\bibitem[\protect\citeauthoryear{{Tazzari} et~al.,}{{Tazzari}
  et~al.}{2016}]{tazzari_2016}
{Tazzari} M.,  et~al., 2016, \mn@doi [\aap] {10.1051/0004-6361/201527423},
  \href {https://ui.adsabs.harvard.edu/#abs/2016A&A...588A..53T} {588, A53}

\bibitem[\protect\citeauthoryear{{Teague} et~al.,}{{Teague}
  et~al.}{2018}]{teague_2018}
{Teague} R.,  et~al., 2018, \mn@doi [\apj] {10.3847/1538-4357/aad80e}, \href
  {http://adsabs.harvard.edu/abs/2018ApJ...864..133T} {864, 133}

\bibitem[\protect\citeauthoryear{{Tscharnuter} \& {Gail}}{{Tscharnuter} \&
  {Gail}}{2007}]{tscharnuter_2007}
{Tscharnuter} W.~M.,  {Gail} H.~P.,  2007, \mn@doi [\aap]
  {10.1051/0004-6361:20065794}, \href
  {https://ui.adsabs.harvard.edu/\#abs/2007A&A...463..369T} {463, 369}

\bibitem[\protect\citeauthoryear{{Visser}, {Doty}  \& {van Dishoeck}}{{Visser}
  et~al.}{2011}]{visser_2011}
{Visser} R.,  {Doty} S.~D.,   {van Dishoeck} E.~F.,  2011, \mn@doi [\aap]
  {10.1051/0004-6361/201117249}, \href
  {http://ukads.nottingham.ac.uk/abs/2011A%26A...534A.132V} {534, A132}

\bibitem[\protect\citeauthoryear{{Walsh}, {Herbst}, {Nomura}, {Millar}  \&
  {Weaver}}{{Walsh} et~al.}{2014}]{walsh_2014}
{Walsh} C.,  {Herbst} E.,  {Nomura} H.,  {Millar} T.~J.,   {Weaver} S.~W.,
  2014, \mn@doi [Faraday Discussions] {10.1039/C3FD00135K}, \href
  {https://ui.adsabs.harvard.edu/\#abs/2014FaDi..168..389W} {168, 389}

\bibitem[\protect\citeauthoryear{{Walsh}, {Nomura}  \& {van Dishoeck}}{{Walsh}
  et~al.}{2015}]{walsh_2015}
{Walsh} C.,  {Nomura} H.,   {van Dishoeck} E.,  2015, \mn@doi [\aap]
  {10.1051/0004-6361/201526751}, \href
  {https://ui.adsabs.harvard.edu/\#abs/2015A&A...582A..88W} {582, A88}

\bibitem[\protect\citeauthoryear{{Weidenschilling}}{{Weidenschilling}}{1977}]{weidenschilling_1977}
{Weidenschilling} S.~J.,  1977, \mnras, \href
  {http://adsabs.harvard.edu/abs/1977MNRAS.180...57W} {180, 57}

\bibitem[\protect\citeauthoryear{{Youdin} \& {Lithwick}}{{Youdin} \&
  {Lithwick}}{2007}]{youdin_2007}
{Youdin} A.~N.,  {Lithwick} Y.,  2007, \mn@doi [\icarus]
  {10.1016/j.icarus.2007.07.012}, \href
  {https://ui.adsabs.harvard.edu/#abs/2007Icar..192..588Y} {192, 588}

\bibitem[\protect\citeauthoryear{{Yu}, {Willacy}, {Dodson-Robinson}, {Turner}
  \& {Evans}}{{Yu} et~al.}{2016}]{yu_2016}
{Yu} M.,  {Willacy} K.,  {Dodson-Robinson} S.~E.,  {Turner} N.~J.,   {Evans}
  Neal~J. I.,  2016, \mn@doi [\apj] {10.3847/0004-637X/822/1/53}, \href
  {https://ui.adsabs.harvard.edu/#abs/2016ApJ...822...53Y} {822, 53}

\bibitem[\protect\citeauthoryear{{Zhang}, {Bergin}, {Blake}, {Cleeves}  \&
  {Schwarz}}{{Zhang} et~al.}{2017}]{zhang_2017}
{Zhang} K.,  {Bergin} E.~A.,  {Blake} G.~A.,  {Cleeves} L.~I.,   {Schwarz}
  K.~R.,  2017, \mn@doi [Nature Astronomy] {10.1038/s41550-017-0130}, \href
  {https://ui.adsabs.harvard.edu/\#abs/2017NatAs...1E.130Z} {1, 0130}

\bibitem[\protect\citeauthoryear{{Zhu}, {Hartmann}, {Nelson}  \&
  {Gammie}}{{Zhu} et~al.}{2012}]{zhu_2012}
{Zhu} Z.,  {Hartmann} L.,  {Nelson} R.~P.,   {Gammie} C.~F.,  2012, \mn@doi
  [\apj] {10.1088/0004-637X/746/1/110}, \href
  {https://ui.adsabs.harvard.edu/#abs/2012ApJ...746..110Z} {746, 110}

\makeatother
\end{thebibliography}




\appendix

\section{Dependence on model parameters}
\label{App:Params}

\begin{figure*}
    \centering
    \includegraphics[width=\textwidth]{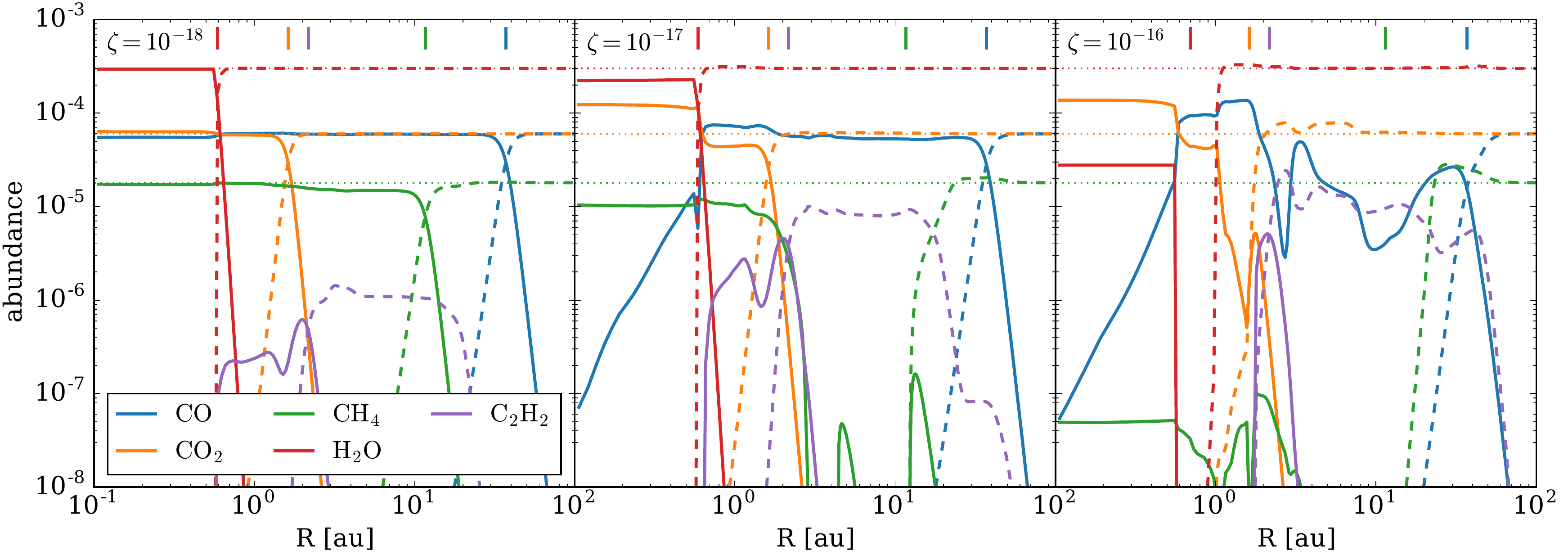} \\
    \includegraphics[width=\textwidth]{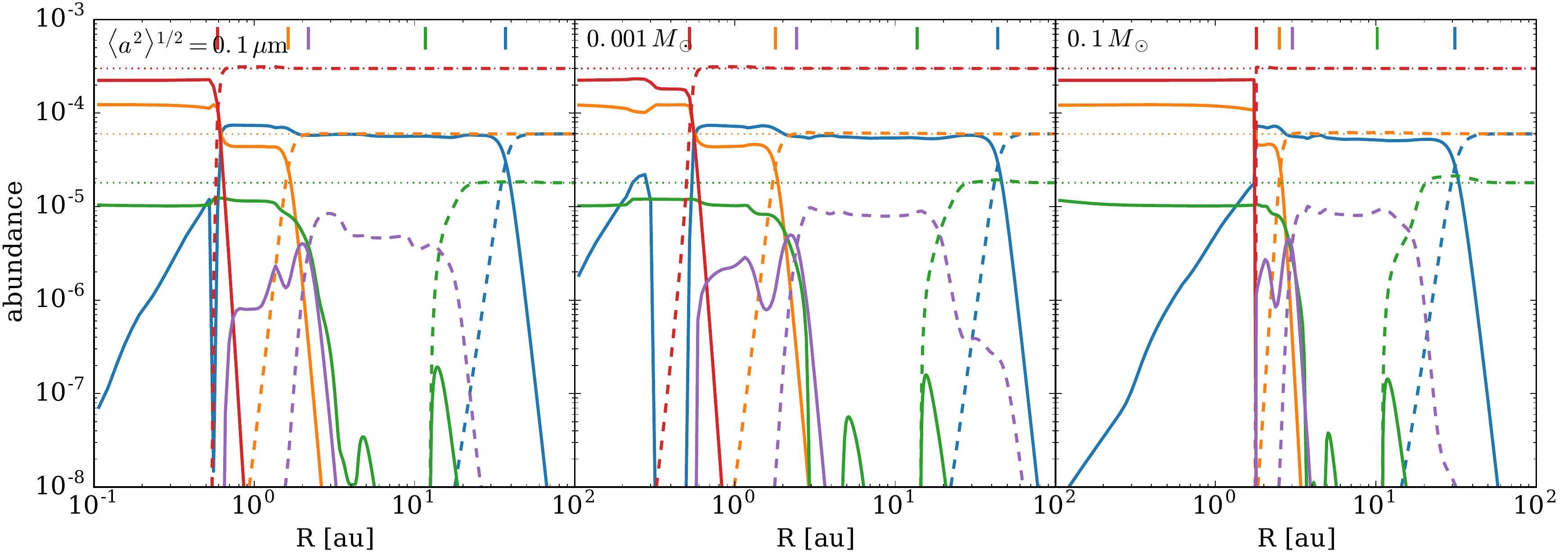}
    \caption{Dependence of the abundance of the main carbon and oxygen carriers on the disc properties in models without transport. Top: Models with varying cosmic-ray ionization rate, but fixed disc mass, $M_{\rm d} = 0.01\,M_\odot$: Bottom Left: A model where the grain size has been fixed at $0.1\,\micron$ in the chemical model. Bottom Middle and Right: Models with varying disc mass.}
    \label{fig:ion_rate}
\end{figure*}

\begin{figure}
    \centering
    \includegraphics[width=\columnwidth]{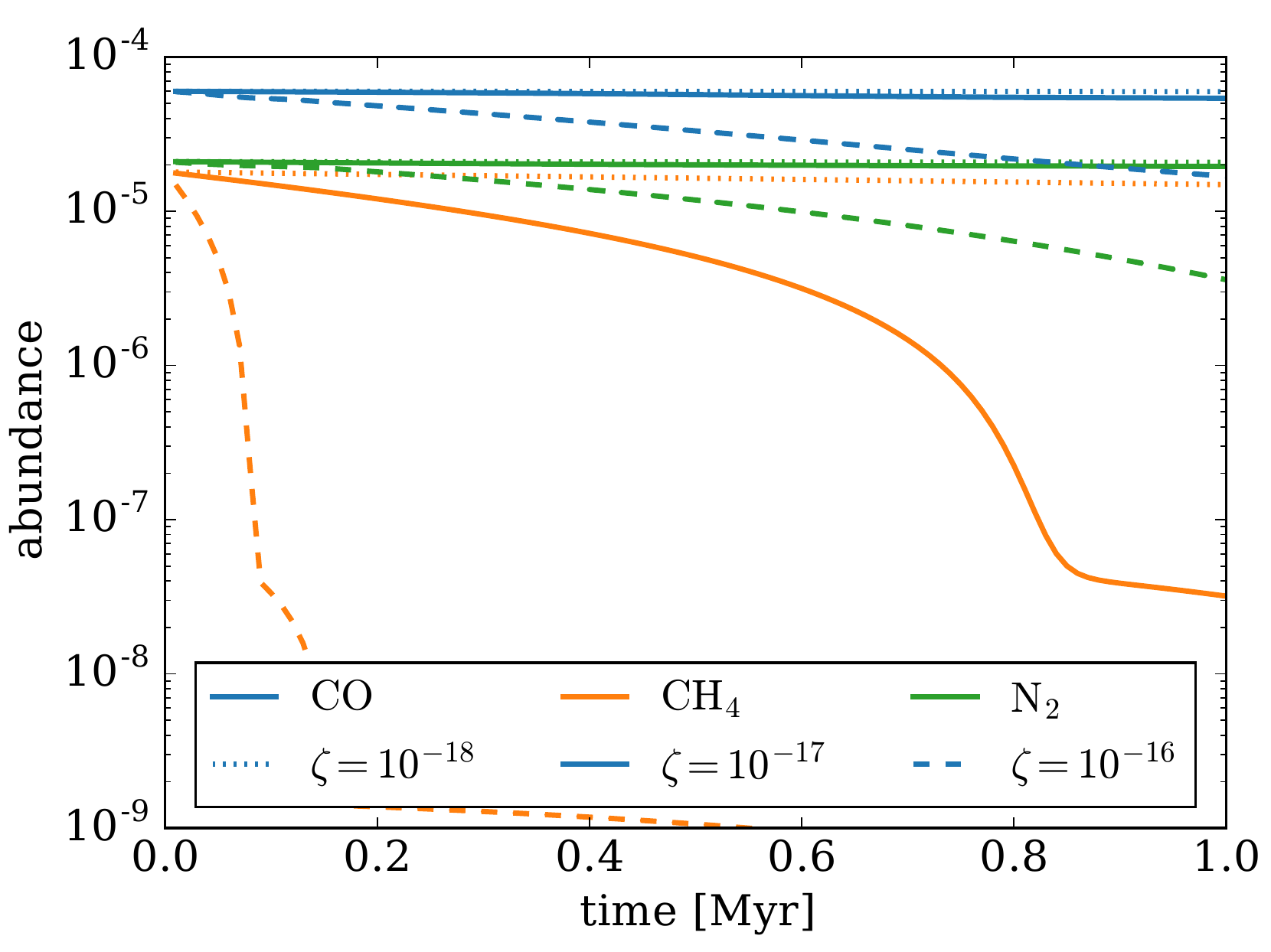}
    \caption{Time evolution of the CO, CH$_4$ and N$_2$ abundance at 5~au in models with varying cosmic-ray ionization rates.}
    \label{fig:ion_evo}
\end{figure}

Here we investigate how the conclusions of our study might depend on the assumed parameters of the model. In particular we vary the cosmic-ray ionization rate, the disc mass, and the average grain size used in the chemical models. For brevity, we report only the results of models neglecting transport. 

The cosmic-ray ionization rate, $\zeta$, is the most important parameter. \autoref{fig:ion_rate} shows that the depletion of CH$_4$ and CO from the gas phase are sensitive to this parameter. Reducing the cosmic-ray ionization rate to $\zeta = 10^{-18}\,s^{-1}$ is sufficient to significantly reduce the depletion of CH$_4$.  Nevertheless, CH$_4$ is still depleted by a factor $\sim 10$ within a Myr, which is fast enough to compete with viscous transport at $\alpha = 10^{-3}$, but not $\alpha = 10^{-2}$. Thus our suggestion that mid-plane chemical kinetics will have a significant impact at low accretions remains true. Even lower ionization rates would slow the depletion of CH$_4$ further. However, short-lived radioactive nuclei are expected to prevent ionization rates falling much below $10^{-18}\,{\rm s}^{-1}$ \citep[e.g.][]{cleeves_2013b}.

Higher cosmic-ray ionization rates would lead to qualitatively different conclusions. At $\alpha = 10^{-3}$ and below, the depletion of CO from the gas phase would now become significant within 1~Myr. Furthermore, the time-scale for the depletion of CH$_4$ reduces to $\sim 0.1\,{\rm Myr}$ (\autoref{fig:ion_evo}), which is now comparable with the viscous transport time-scale at $\alpha = 10^{-2}$. Under such conditions transport is unlikely to strongly suppress the depletion of CO and CH$_4$. However, we note that such high ionization rates are not typically expected for T Tauri star discs, instead the cosmic-ray ionization rates are mpre likely lower than the canonical value of $10^{-17}\,{\rm s}^{-1}$ \citep{cleeves_2013}.

In addition to varying the ionization rate, we conducted models in which the disc mass has been varied. We find that the surface density in the disc does not greatly affect the composition, as can be seen by comparing the models with $M_{\rm d} = 0.001\,M_\odot$ and $0.01M_\odot$. At $M_{\rm d} = 0.1\,M_\odot$ we see differences, but these are due to the higher temperature in the disc, rather than higher density. Since these differences are limited to within a few au from the star where the transport time-scales are short, we conclude that disc mass does not much affect our conclusions.

Finally, we tested the effects of our assumptions about the grain size distribution. In the main text we compute the average grain area assuming a grain size distribution with $n(a) {\rm d}a \propto a^{-3.5} {\rm d}a $ between 0.1$\micron$ and $a_{\rm max}$, where $a_{\rm max}$ is set by the dust evolution model. To test this assumption we compare our results with a model where the grain size used in the gas-grain reactions is fixed at $0.1\micron$. \autoref{fig:ion_rate} shows that the evolution is largely insensitive to the average grain area.


\bsp	
\label{lastpage}
\end{document}